\documentclass[aps,roupedaddress,draft,showpacs,showkeys]{revtex4-1}

\usepackage{amssymb}
\usepackage{amsmath}

\begin{document}

%\preprint{}

\title{A Process Model of Quantum Mechanics}


\author{William H. Sulis}
\affiliation{McMaster University and The University of Waterloo}
\email{sulisw@mcmaster.ca}

\date{\today}

\begin{abstract}
A process model of quantum mechanics utilizes a combinatorial game to generate a discrete and finite causal space upon which can be defined a self-consistent quantum mechanics. An emergent space-time $\mathcal{M}$ and continuous wave function arise through a non-uniform interpolation process. Standard non-relativistic quantum mechanics emerges under the limit of infinite information (the causal space grows to infinity) and infinitesimal scale (the separation between points goes to zero). This model has the potential to address several paradoxes in quantum mechanics while remaining computationally powerful.
\end{abstract}


\maketitle
\section{Introduction}

Questions concerning the completeness of quantum mechanics date back at least to the famous paper of Einstein, Podolsky and Rosen \cite{Einstein}. Subsequent 
research into hidden variable theories \cite{Hemmick} and quantum information \cite{Barrett} has led to a series of results that rather strongly suggest that quantum mechanics is a complete theory, at least in so far as its probabilistic and statistical structure is concerned. A different approach based on a model of process \cite{Whitehead, Shimony} developed within a complex systems theory framework \cite{Sulisad, Sulisct, Sulisjmp} suggests that the answer might not be so clear cut, and that quantum mechanics, at least non-relativistic quantum mechanics (NRQM), might be missing important dynamical information. Moreover, there have also been long debates about whether reality is local or non-local \cite{Maudlin1,Bohm1,Aharonovp, Greenstein,Larsson}, contextual or non-contextual \cite{Kochen, Leggett, Mermin, Pusey}, whether the underlying probability structure is Kolmogorov or non-Kolmogorov \cite{Gleason,Khrennikov,Khrennikov1,vonNeumann} and whether the wave function is real or merely heuristic \cite{Colbeck2,Ney, Spekkens}. Several recent papers have attempted to prove from a quantum information perspective that the wave function must correspond to something physical and not merely represent knowledge \cite{Barrett1,Colbeck1,Ord}. 

This paper aims to contribute to these debates. It presents a decidedly unromantic (to quote Bell \cite{Bell}) model of quantum mechanics, grounded in process theory (see Shimony \cite{Shimony}), in which wave functions correspond to real physical waves, in which space-time and physical entities are emergent, and which is discrete, finite, intuitive, causal, quasi-local and quasi-non-contextual, yet retains the computational power of standard quantum mechanics. The model considers a discrete, finite causal space which is \emph{generated} rather than simply existing. The physics plays out entirely on this causal space using only causally local information. Non-relativistic quantum mechanics appears as an idealization when information can be considered to be infinite and the spacing between elements of the causal space to be infinitesimal.

There have been previous attempts to provide a realist, emergent, or process based quantum mechanics: Bohmian mechanics \cite{Bohm2}, Wolfram's cellular automata \cite{Wolfram}, continuous spontaneous localization (CSL) \cite{Ghirardi}, Finkelstein's quantum relativity \cite{Finkelstein1,Finkelstein}, Noyes's bit-string physics \cite{Noyes}, Bastin and Kilmister's combinatorial physics \cite{Bastin}, Hiley's process physics \cite{Hiley}, Cahill's process physics \cite{Cahill}, Bodiyono's fluctuation model \cite{Bodiyono, Bodiyono1, Bodiyono2, Bodiyono3}.  There have been attempts to formulate quantum mechanics using different mathematics frameworks such as quantum information \cite{Chiribella}, quantum logic \cite{Beltrametti} and category theory \cite{Coecke}. Bohmian mechanics requires an extreme non-locality in the form of the quantum potential. CSL requires repeated wave function collapse while still retaining the original wave function as a guiding influence. None of the others have gained much purchase within the larger physics community, in part perhaps because the path back to standard quantum mechanics is not straightforward.

The process approach promises to place quantum mechanics on a solid realist foundation, with a minimum of non-locality and contextuality, free of paradox, free of divergence, while still retaining computational power.

The idea of process originates with Whitehead \cite{Whitehead} and has been explored previously in physics \cite{Shimony, Eastman} but never fully 
developed. The process model presented here has its origins in efforts to study information flow in complex systems such as in brains \cite{Sulist, Sulistn}, collective intelligence \cite{Sulisci}, dynamical networks \cite{Trofimova}, gene-regulatory networks, and economic systems. The idea of a discrete causal space was inspired by Sorkin's causal set programme \cite{Sorkin} but diverges from it in its implementation. The necessity to consider discrete models is suggested by the existence of wave-particle duality, by considerations in quantum gravity, by the need to avoid divergences in quantum field theories, and by recent work by Gisin \cite{Gisin}, who constructed a Bell type inequality showing that either one must reject the principle of continuity or accept instantaneous information transfer between space-like separated entities (thus rejecting special relativity). In the model presented below, continuity is recovered via an interpolation procedure (inspired by an idea of Kempf \cite{Kempf}) and quantum mechanics (at least non-relativistic quantum mechanics) arises directly as an effective theory when a continuum idealization can be assumed. 

\section{Informons}

The key idea, based on Whitehead's Process Theory \cite{Eastman} assumes that the elements of physical reality do not simply \emph{exist}, but rather are \emph{emergent} upon a lower level of entities. Whitehead called these entities \textquotedblleft actual occasions\textquotedblright \, but they shall be referred to as \emph{informons} (short for informational monads) to reflect their fundamental informational character. These informons are generated moment to moment through the actions of processes, which interpret, transform and supplement the information of the current generation of informons and incorporate it into the next.
Informons are postulated to be discrete, delocalized, finite, and organized into distinct generations. They are transient - arising, persisting briefly as their information is incorporated into the next generation of informons, then abating, much like the pixels on a computer screen. Physical entities appear at the emergent level analogous to the image patterns that form on the screen. Information passes causally from one generation to the next, \emph{never} within a generation. As a consequence, special relativity is not violated \cite{Sulis}. 

Space-time, and the physical entities that manifest in space-time, are all postulated to be emergent from these informons which precludes any possibility of direct observation by the physical entities that emerge from them. Ironically, informons must be \emph{interpreted} in reference to the entities they generate \cite{Sulis1}. This is because observations of informons can only occur through an act of \emph{measurement}, which is understood from the process perspective to involve an interaction between a system process and a dynamically specialized measurement apparatus process.  

Each informon $n$ is assigned a tuple $\mathbf{p}_{n}$ of properties inherited from the process $\mathbb{P}$ that generates it. The most important of these is the \textquotedblleft strength \textquotedblright\, or \textquotedblleft coupling effectiveness \textquotedblright\, $\Gamma_{n}$ of the generating process $\mathbb{P}$ as determined at the informon $n$. Here, coupling refers to interactions between $\mathbb{P}$ and other processes. The coupling effectiveness expresses the compatibility between informons of different processes and therefore their likelihood of interacting. This notion of compatibility arises in complex system theory and was first proposed by Trofimova \cite{Trofimova} in her work on ensembles with variables structures. As will be seen below, the strength $\Gamma(n)$ enters into the construction of the wave function but it does \emph{not} determine the probability of occurrence of informons, rather it determines the probability of that informon triggering off the registration of a measurement when in interaction with a measurement process. The probability associated with measurements thus becomes another emergent feature of the process viewpoint.

Although previous generations fade from existence their information propagates to subsequent generations in a manner akin to a dissipative wave. Following an idea of Markopoulou \cite{Markopoulou}, this prior information is associated with the current informon in the form of a content set $G_{n}$. $G_{n}$ then consists of all of those informons from previous generations that pass information to $n$. The passing of information from one generation to another induces a causal structure between generations. Given two informons $m,m'$ in $G_{n}$, write $m\rightarrow m'$ (or $m<m'$) if $m$ was in a earlier generation than $m'$ and information was passed from $m$ to $m'$.  $G_{n}$ is thus an acyclic directed graph whose vertices consist of these prior informons. Equivalently it may be viewed as a partially ordered set. Note that $m<n$ for every $m\in G_{n}$. Furthermore, if $\mathcal{I}$ represents a single generation of informons, then $\mathcal{I}$ forms an antichain and $\cup_{n\in \mathcal{I}} G_{n}$ also forms an acyclic directed graph, ensuring consistency of the causal structure. 

One generation $\mathcal{I}$ of informons represents an instance of reality, with the physics playing out in the succession of generations. The information upon which the process dynamics depends must therefore lie within $\{\mathbf{p}_{n}|n\in \mathcal{I}\}$ and $\{G_{n}|n\in \mathcal{I}\}$.

In many cases it is also useful to assume that there is an indefinite metric $d$ assigned to $L=\mathcal{I} \cup_{n\in \mathcal{I}} G_{n}$ such that for any $m,m'\in K$, $d(m,m')>0$ iff $m<m'$. This assigns a \textquotedblleft time-like\textquotedblright\, distance between successive generations and a \textquotedblleft space-like\textquotedblright\, distance within a generation and between informons of distinct generations that are causally unrelated.

Although informons are transient, it is useful to exploit the artifice of a history of their appearance. If $\mathcal{I}_{n}$ is the current generation, let $\mathcal{I}_{<n}=\cup_{k}\mathcal{I}_{k}$, $k<n$ (note we allow $k<0$).

Informons are postulated to be observable only through measurement. Thus they can at best be interpreted through some physical theory and associated with elements of some related model. Since the concern here is with quantum mechanics, the models of interest are causal manifolds (a causal manifold is a generalization of the light cone structure of Minkowski space to arbitrary manifolds \cite{Borchers}) and Hilbert spaces over such manifolds. Although in general there may be a myriad of ways in which these informons may be related to elements of these causal spaces and Hilbert spaces, there should be some limitations to ensure that at least the statistical behaviour of the resulting interpretation does not depend upon the exact details.
 
Each informon $n\in \mathcal{I}_{k}$ is therefore interpreted as a point $\mathbf{x}_{n}$ in a causal manifold $\mathcal{M}$ with causal ordering $\prec$ and metric $\rho$. Given $n,n'$, if $n<n'$ then $\mathbf{x}_{n}\prec \mathbf{x}_{n'}$. Sometimes one requires that $d(n,n')=\rho(n,n')$ or that $\rho(n,n')\leq d(n,n')\pm\epsilon$ for a small error $\epsilon$. The wave function $\Psi(\mathbf{z})\in\mathcal{H}(\mathcal{M})$ of a physical entity (including the vacuum) is understood to represent  a physical wave defined on the causal manifold $\mathcal{M}$. Each informon is interpreted as providing a local Hilbert space contribution to this wave function of the form $\phi_{n}(\mathbf{z})=\Gamma_{n}f_{n}(\mathbf{z},\mathbf{x}_{n})$, such that $\Psi(\mathbf{z})=\sum_{n\in \mathcal{I}_{k}}\phi_{n}(\mathbf{z})$. Where definable, $f_{n}$ will be a translation of a single generating function $g$, that is $f_{n}(\mathbf{z},\mathbf{x}_{n})=T_{\mathbf{x}_{n}}g(\mathbf{z})=g(\mathbf{z}-\mathbf{x}_{n})$.
The pair $(\mathbf{x}_{n},\phi_{n}(\mathbf{z}))$ is called the \emph{interpretation} of $n$, $\mathbf{x}_{n}$ the local $\mathcal{M}$-interpretation and $\phi_{n}(\mathbf{z})$ the local $\mathcal{H}(\mathcal{M})$-interpretation. It is essential to understand that the interpretation is imposed upon the informon by an external observer and is not an intrinsic feature of the informon itself. The physics on the causal space is self-contained and does not depend upon the interpretation. It does, however, induce an apparent dynamic upon the interpretation which is observable. 

Given this understanding, an informon will be denoted simply as $[n]<\alpha_{n}>\{G_{n}\}$ where $\alpha_{n} = (\mathbf{x}_{n},\phi_{n}(\mathbf{z}),\Gamma_{n},\mathbf{p}_{n})$. A generation $\mathcal{I}$ of such informons is also referred to as a causal tapestry \cite{Sulis}

\section{Process}

Informons are postulated to be generated by processes. Processes possess only algebraic properties: they generate space-time and so cannot be situated in space-time. A process may be active, in which case it acts in a series of rounds to generate informons, or inactive. Each process $\mathbb{P}$ is described by several parameters: one or more tuples of properties ($\mathbf{p}\in D$) (such as mass, charge, spin, energy, angular momentum, linear momentum, lepton number etc), the number of informons generated during a round ($R$), the number of previous informons whose information is incorporated into a nascent informon ($r$), the number of rounds needed for a generation ($N$), the temporal and spatial scales of informons ($t_{P},l_{P}$). These scales could be universal, applying to all physical entities simultaneously, or they could be individual, applying to each entity separately and linked to the energy ($E\propto 1/t_{P}$) and momentum ($p\propto 1/l_{P}$) of the entity. It is left for future experimentation to determine which assumption is more consistent with observation. 

A process creates a single generation of informons and the history of a system comprises a succession of such generations.

A system unfolds as a succession of such present moments.

\begin{displaymath}
\cdots\mathcal{I}_{n}\stackrel{\mathbb{P}_{n}}{\rightarrow}\mathcal{I}_{n+1}\stackrel{\mathbb{P}_{n+1}}{\rightarrow}\mathcal{I}_{n+2}\cdots
\end{displaymath}

\noindent The triple of current generation $\mathcal{I}_{k}$, process $\mathbb{P}$, and nascent generation $\mathcal{I}_{k+1}$, forms what philosophers term a compound present.

A \emph{primitive} process is defined as generating a single informon during a single round ($R=1$). Intuitively, the action of a primitive process is to generate, one by one, a succession of informons $n_{1},n_{2},\ldots $, thus forming a generation $\mathcal{I}_{n}$. The sequential generation of informons by a process is a key feature of the model. It is essential that informons be generated sequentially to ensure that interactions between processes, especially with measurement processes, be triggered by single informons and thereby exhibit both quantization as well as determinate properties. Measurements are triggered by single informons, resulting in a discrete transfer of information, including that concerning properties such as energy and momentum, and thus imparting a particle-like quality to interactions. At the emergent level, however, the process generates a discrete sampling of a physical wave, thus imparting a wave-like quality to its behaviour. Physical entities from the process perspective possess both particle and wave properties depending upon the scale. 

Each generation $\mathcal{I}_{n}$ forms a discrete space-like slice of space-time, and being an antichain, embeds into $\mathcal{M}$ as a discrete sampling $\{\mathbf{x}_{n_{i}}\}$ of a space-like hyper-surface $M$. They also form a discrete sampling $\{\phi_{n_{i}}(\mathbf{z})\}$ of a wave function $\Psi(\mathbf{z})$ defined on that surface by $\Psi(\mathbf{z})\approx\sum_{i}\phi_{n_{i}}(\mathbf{z})$ (actually $\Psi(\mathbf{z})$ is defined on $\mathcal{M}$ but the approximation is best when restricted to the hyper-surface). The coupling factor at $n_{i}$ is defined as $\Gamma_{n_{i}}=\phi_{n_{i}}(\mathbf{m}_{n_{i}})$ and the strength there is defined as $l_{P}^{3}\Gamma_{n_{i}}^{*}\Gamma_{n_{i}}$. The relative strength of $\mathbb{P}$ over the causal space $\mathcal{I}$, $||\mathbb{P}||_{\mathcal{I}}$ is defined as $||\mathbb{P}||_{\mathcal{I}}^{2}=\sum_{n_{i}\in \mathcal{I}}l_{P}^{3}\Gamma_{n_{i}}^{*}\Gamma_{n_{i}}$. Applying the same interpretations to the complete history $\mathcal{I}_{\infty}$ yields a discrete version of $\mathcal{M}$ and a discrete sampling  of a global the wave function on $\mathcal{M}$. Standard quantum mechanics can be viewed as an idealization or as an effective theory in the limit of infinite information ($r,N\rightarrow \infty$) and sometimes infinitesimal spacing ($t_{P},l_{P}\rightarrow 0$). The conditions under which this is possible depend upon the form of the generating function $g$ for the local Hilbert space contribution and the geometry of the embedding into $\mathcal{M}$, and the effectiveness of the interpolation can be determined from various interpolation theories \cite{Zayed,Jorgensen}. 

The relationship between process and time is open to multiple interpretations and is worthy of deeper philosophical analysis. Process may be viewed as occurring \emph{outside} of space-time, generating a causal space which can be interpreted as a 4-dimensional causal manifold, one dimension of which the observer associates with time. Process may equally well be considered \emph{as} time, as expressing the arrow of time, and again generating a causal space interpretable as a 4-dimensional causal manifold where the observer associates one dimension with this underlying time. Time could be the usual time, with processes generating the causal space on a near infinitesimal time scale, so that to the observer the generation of the causal space appears instantaneous. Another possibility is that process occurs following a separate, second time, akin to that suggested in the two-time physics of Bars \cite{Bars} or in the stochastic quantization of Parisi and Wu \cite{Parisi}. Regardless, it is suggested that the compound present formed by the triple $\cdots\mathcal{I}_{n}\stackrel{\mathbb{P}_{n}}{\rightarrow}\mathcal{I}_{n+1}$ forms the basis for the personal experience of time and of a moving present. Future research may suggest which, if any, of these possibilities is most reasonable.

Processes are generally considered to act non-deterministically, a term used in computation theory to mean that actions are described by set-valued maps without any intrinsic probability structure. Probabilities arise through two mechanisms: combinatorial proliferation, similar to the case for iterated function systems, and coupling of processes, in particular through couplings to measurement processes. Of course one can consider deterministic or stochastic processes but these are of less interest. One cannot do justice to this topic in a short note and the theory of interaction and measurement provided below is necessarily brief and will be discussed in more detail in a separate letter. 

In the case that the Lagrangian uses generalized coordinates there may be subtleties in the interpolation procedure depending upon the geometry of these coordinates but that would take this discussion too far afield. It is also important to note that when process is represented by means of combinatorial games as described below, the process covering map will become dependent not only upon the particular interpolation scheme utilized but also upon the particular strategy used to implement the game.

Conservation laws and symmetries applied to the properties of processes provide a set of algebraic constraints upon possible interactions among processes. These are inherited by the wave functions through the process covering map suggesting that processes are primary and wave functions secondary. Quantum mechanics may be best viewed as an effective theory, valid under certain asymptotic limits, but not necessarily the final theory. 

\section{Interactions and the Algebra of Process}

The generation of an informon can trigger a coupling between processes or the activation or inactivation of processes, depending upon the compatibility of these informons \cite{Trofimova1}. Couplings between processes take many forms, providing the space of processes with a rich algebraic and combinatorial structure. Processes may act sequentially (denoted as sums) or concurrently (denoted as products). They may act independently of one another (independent) or their actions may be constrained, so that the action of one forces limitations on the actions of another (interactive). They may act on the same nascent informon (free or bosonic) or only on distinct informons (exclusive or fermionic). These consideration give rise to 8 general possibilities - a) Sequential sums: $\hat\oplus$ (free, independent), $\oplus$ (exclusive, independent), $\hat\boxplus$ (free, interactive), $\boxplus$ (exclusive, interactive), b) Concurrent products: $\hat\otimes$ (free, independent), $\otimes$ (exclusive, independent), $\hat\boxtimes$ (free, interactive), $\boxtimes$ (exclusive, interactive). The interactive case is actually shorthand for a set of possible interactions. 

The character of a process is determined mostly by the types of its properties while different states are determined by their values. As a rough rule of thumb for considering interactions, like processes may sum or multiply, but unlike processes only multiply. Primitive processes multiply to generate multiple informons during a round ($R>1$). Complex processes are formed through algebraic combinations of simple processes. Processes corresponding to different states of a single entity combine through the exclusive sum. The reason for this is to ensure that any single informon representing an instance of the entity corresponds to just a single assignment of properties, thereby assigning a definite reality to the informon. Processes may superimpose but actual occasions corresponding to a single entity \emph{never} superimpose. However, an actual occasion may form an instance of multiple distinct processes, and therefore multiple distinct physical entities. Subdivision of a process based upon boundary conditions, as in the case of the two slit experiment described below, utilizes the free sum. This is because a division of a process via boundary conditions does does not alter any intrinsic properties of the process, and so each contribution is merely a restricted action of the same process.

A superposition of simple processes corresponding to a single physical entity thus takes the form $\oplus_{i} \mathbb{P}_{i}$, meaning that during a given round only \emph{one} of the $\mathbb{P}_{i}$ is active, generating a single informon. The choice of subprocess may change from round to round, but any informon is generated \emph{only} by a single subprocess. This interleaving of subprocesses will, through the process covering map described below, nevertheless approximate the standard wave function for the superposition. More generally one has $\oplus_{i} w_{i}\mathbb{P}_{i}$, where $w_{i}\mathbb{P}_{i}$ indicates a modification of certain attributes of $\mathbb{P}_{i}$ by the factor $w_{i}$, for example multiplying the wave function contribution by $w_{i}$. 

If $\mathbb{P}$ is a primitive process generating a single entity, then a process generating $N$ such independent entities could be given by either $\overbrace{\mathbb{P}\hat\otimes\cdots \hat\otimes\mathbb{P}}^{N}$ (bosonic) or $\overbrace{\mathbb{P}\otimes\cdots \otimes\mathbb{P}}^{N}$ (fermionic).

Entanglement provides an example of an interactive coupling between processes. An entanglement of primitive processes is denoted as $\boxtimes_{i} \mathbb{P}_{i}$, (or $\hat\boxtimes_{i} \mathbb{P}_{i}$) meaning that during a given round, \emph{all} of the $\mathbb{P}_{i}$ are concurrently generating single informons, but because they are in interaction mode their actions are mutually constrained, and the resulting informons will be correlated. This is more apparent when processes are represented as combinatorial games where the game tree of an interactive product will be a proper subtree of the game tree of the independent products. Note that no information passes among these informons. Each is generated by its own process using only causally local information propagating from current informons linked to that process. There is no \textquotedblleft spooky action at a distance\textquotedblright. There is just the realization that the generating sub-processes are not independent of one another but rather are in an interactive mode.

The zero process, $\mathbb{O}$, is the process that does nothing. It generates no informons at all. It is quite possible that  certain processes cannot be combined. This might occur because of superselection rules, or because as primitive processes are combined in ever more complex ways configurations may be generated that violate algebraic constraints. For example, it is not reasonable to believe that processes having different characters of strength (scalar, vector, spinorial) could be combined. In such a case a sum of such incompatible processes yields the zero process. For example, a scalar process $\mathbb{P}_{s}$ and a spinorial process $\mathbb{P}_{sp}$ form the sum $\mathbb{P}_{s}\oplus \mathbb{P}_{sp}=\mathbb{O}$. Similarly there may be situations in which it is impossible to form the product of two or more distinct processes, in which case the product will be given the value $\mathbb{O}$. 

It is conjectured that the emergence of classicality derives not merely from statistical effects or from idealizations in which $\hslash\rightarrow 0$ but rather as the number of participating components increases, leading to the appearance of non-superposable processes, particularly when combined in interactive sums and products. By this is meant the appearance of distinct processes $\mathbb{P}_{a},\mathbb{P}_{d}$ that could, in principle, generate the evolution of a system separately but cannot be superposed so as to generate an evolution in combination, i.e. $\mathbb{P}_{a}\oplus\mathbb{P}_{d}=0$. 
Processes may also be concatenated. Given a sequence of active processes 
\begin{displaymath}
\mathcal{I}_{n}\stackrel{\mathbb{P}_{n}}{\rightarrow}\mathcal{I}_{n+1}\stackrel{\mathbb{P}_{n+1}}{\rightarrow}\mathcal{I}_{n+2}
\end{displaymath}

one may form the concatenated process

\begin{displaymath}
\mathcal{I}_{n}\stackrel{\mathbb{P}_{n}\mathbb{P}_{n+1}}{\rightarrow}\mathcal{I}_{n+2}
\end{displaymath}

In general concatenations are non-commutative.

\section{Process Covering Map}
 
The process viewpoint leads to the insight that the proper setting for quantum dynamics is not the Hilbert space $\mathcal{H}(\mathcal{M})$ but rather the space $\Pi$ of generating processes. Dynamical information is lost when considering only the wave function. In a process driven world, informons are generated and since interactions such as measurement are held to be triggered by the appearance of informons (which determine the couplings of processes), the sequence of appearance of informons during the unfolding of processes potentially carries dynamically relevant information. Moreover, processes are determined not just by their parameters but also by the strategy by means of which informons are generated. This will be made more apparent when the game representation is discussed below. The process strategy determines the sequence of generation of informons, their causal distances, their properties and their strength. The generation  sequence potentially conveys information about the generation strategy.

Beginning with some generation $\mathcal{I}$ of informons, a primitive process $\mathbb{P}$, once activated, will generate a sequence of $n_{1}, n_{2}, \ldots, n_{k}, \ldots$ of informons. Due to the non-deterministic nature of the process action, reactivating the process on the same initial generation will, in all likelihood, result in a different sequence of informons, say $m_{1}, m_{2}, \ldots, m_{k}, \ldots$. The sequence tree of a process $\mathbb{P}$ and initial generation $\mathcal{I}$, denoted $\Sigma(\mathbb{P},\mathcal{I})$ is constructed as follows. Into level $0$ put the empty informon $\emptyset$. Into level $1$, place all possible informons that can be generated by $\mathbb{P}$ starting from $\mathcal{I}$ in the first round. Form a link between $\emptyset$ and each of these informons. From each $n_{1}$ in level $1$ form a link to all possible informons that can be generated following $\mathcal{I}$, $n_{1}$, and form a link between them and $n_{1}$. Level $2$ consists of all such informons generated from all informons in level $1$.  This construction process is repeated for all possible generations. Any actual implementation of the process $\mathbb{P}$ will generate one sequence which constitutes a path through the sequence tree. 

The informons $n_{1}, n_{2}, \ldots, n_{k}, \ldots$ along each path of $\Sigma(\mathbb{P},\mathcal{I})$ constitute a possible next generation of informons. Each informon $n_{i}$ will provide a local $\mathcal{H}(\mathcal{M})$-contribution $\phi_{n_{i}}(\mathbf{z})$ to a global $\mathcal{H}(\mathcal{M})$-interpretation $\Phi^{1}(\mathbf{z})=\sum_{n_{i}} \phi_{n_{i}}(\mathbf{z})$. Thus to the process $\mathbb{P}$ one can associate a set $H_{\mathbb{P}}$ of elements of $\mathcal{H}(\mathcal{M}))$ consisting of all global $\mathcal{H}(\mathcal{M})$-interpretations constructed from every possible path in the sequence tree $\Sigma(\mathbb{P},\mathcal{I})$.
 
Interpolation theory shows that given certain choices of the interpolation function $g$, in the limit $N,r\rightarrow \infty$, $H(\mathbb{P})\rightarrow \{\Phi^{t_{P}l_{P}}(\mathbf{z})\}$, a single function. Given two distinct primitive processes $\mathbb{P}_{1}$ and $\mathbb{P}_{2}$, it is possible that in the limit  $N,r\rightarrow \infty$, $H(\mathbb{P}_{1})\rightarrow H(\mathbb{P}_{2})=\{\Phi^{t_{P}l_{P}}(\mathbf{z})\}$. In such a case the primitive processes $\mathbb{P}_{1},\mathbb{P}_{2}$ are said to be $\Psi$-epistemic equivalent (sometimes weak epistemic equivalent). $\Psi$-epistemic equivalent processes generate distinct \textquotedblleft realities\textquotedblright\, at small scales which are effectively equivalent at large scales, at least statistically. They are indistinguishable from a quantum mechanical point of view since they will asymptotically yield the same wave function. Interpolation theory shows that this asymptotic behaviour holds for a single primitive process  so that $\mathbb{P}$ is $\Psi$-epistemic equivalent to itself. $\Psi$-epistemic equivalence thus provides a proper equivalence relation on the subset of primitive processes. $\Psi$-epistemic equivalence coarse grains the space of processes into those that are equivalent quantum mechanically, which narrows the range of strategies to be considered. This imposes a form of strategic \textquotedblleft gauge\textquotedblright\, invariance, so that the strategies considered in models may be chosen for computational or analytical convenience rather than ontological implications. It is always possible that future advances in measurement might permit strategies within $\Psi$-equivalent classes to be distinguished but that is not the case at present.

The current state of interpolation theory applies mostly to complex vector valued functions while the theory as applied to spinors is much less developed. Thus from a technical standpoint the discussion to follow applies to integral spin particles but it seems reasonable to believe that the argument should  hold for half integer spin particles as well. The most general case requires consideration of general function spaces mapping $\mathcal{M}$ to $\mathbb{C}^{n}$ or $\mathfrak{s}\mathfrak{p}(n)$ but this would unnecessarily complicate the essential points.

The process covering map (PCM) provides a linkage between the space of processes, $\Pi$ and the Hilbert space $\mathcal{H}(\mathcal{M})$, and thus to NRQM. In general the PCM will depend upon the strategy used to implement the actions of the processes as well as the initial condition. For simplicity only the initial condition will be referenced. It is constructed first on the set of primitive processes $\Pi_{p}$. For fixed $\mathcal{I}$, and some primitive process $\mathbb{P}$, define the process covering map $\mathfrak{P}_{\mathcal{I}}:\Pi_{p}\rightarrow \mathcal{H}(\mathcal{M})$  by $\mathfrak{P}_{\mathcal{I}}(\mathbb{P})=H_{\mathbb{P}}$. Note that the PCM is a \emph{set}-valued map.

As will be made explicit in the example below, if the strength $\Gamma(n)\propto \Psi(\mathbf{m}_{n})$ for some non-relativistic wave function $\Psi(\mathbf{z})$ which has energy and momenta bounded away from Planck energy and momentum, then $\Psi(\mathbf{z})=\Psi^{t_{P}l_{P}}(\mathbf{z})$, and if $\Psi(\mathbf{z})$ is not band-limited in this way, then this will still be true in the limit $t_{P},l_{P}\rightarrow 0$. 

The PCM is now extended to $\Pi$ by considering sums and products. 
The case of exclusive sums is easily treated. Consider an exclusive sum $\oplus w_{i}\mathbb{P}_{i}$ of primitive processes $\mathbb{P}_{i}$ applied to the causal tapestry $\mathcal{I}$. The elements of the nascent causal space $\mathcal{I}_{1}$ lie in distinct subsets, $\mathcal{I}_{1}^{i}$. Let $j_{n}=i$ iff $n\in \mathcal{I}_{1}^{i}$. Then the global $\mathcal{H}(\mathcal{M})$-interpretation $\Phi(\mathbf{z})=\sum_{n\in \mathcal{I}_{1}} w_{j_{n}}\phi_{n}(\mathbf{z})=\sum_{i}w_{i}\{\sum_{n\in \mathcal{I}_{1}^{i}}\phi_{n}(\mathbf{z})\}=\sum_{i}w_{i}\Phi^{i}(\mathbf{z})$ where $\Phi^{i}(\mathbf{z})$ is the global $\mathcal{H}(\mathcal{M})$-interpretation corresponding to the process $\mathbb{P}_{i}$. Interpolation theory provides conditions under which, in the asymptotic limit

\begin{displaymath}
\Phi(\mathbf{z})=\sum_{i}w_{i}\Phi^{i}(\mathbf{z})\rightarrow \Psi(\mathbf{z})=\sum_{i}w_{i}\Psi^{i}(\mathbf{z})
\end{displaymath}

It follows easily that $\mathfrak{P}(\oplus_{i} w_{i}\mathbb{P}_{i})=\sum_{i} w_{i}\mathfrak{P}(\mathbb{P}_{i})$ where for two sets of functions $A,B$ the sum $A+B=\{f+g|f\in A,g\in B\}$.

This also holds true for free sums, so that $\mathfrak{P}(\hat\oplus_{i} w_{i}\mathbb{P}_{i})=\sum_{i} w_{i}\mathfrak{P}(\mathbb{P}_{i})$.

It is here that the process point of view departs subtly from standard quantum mechanics. It is easy to see that under the limit $N,r\rightarrow \infty$,
$\mathfrak{P}(\oplus_{i} w_{i}\mathbb{P}_{i})\rightarrow \mathfrak{P}(\hat\oplus_{i} w_{i}\mathbb{P}_{i})$, so that $\oplus_{i} w_{i}\mathbb{P}_{i}$ and $\hat\oplus_{i} w_{i}\mathbb{P}_{i}$ are $\Psi$-epistemic equivalent, but the processes $\oplus_{i} w_{i}\mathbb{P}_{i}$ and $\hat\oplus_{i} w_{i}\mathbb{P}_{i}$ do not necessarily possess the same sequence trees. Moreover the local $\mathcal{H}(\mathcal{M}))$-contributions which determine the coupling to other processes will differ (since in the free case it is possible for an informon to carry contributions from, two distinct sub-processes, which could result in measurements that differ from the exclusive case.  

The case of products is more complicated. Consider the exclusive product $\mathbb{P}=\otimes \mathbb{P}_{i}$ of primitive processes $\mathbb{P}_{i}$. An informon $n^{i}$ corresponding to each $i$ is generated during each round. Therefore a local Hilbert space  contribution $\phi_{n^{i}}$ is generated corresponding to each $i$. In the exclusive case one may again form a global $\mathcal{H}(\mathcal{M})$-interpretation $\Phi(\mathbf{z})=\sum_{n^{i}}\phi_{n^{i}}(\mathbf{z})=\sum_{i}\Phi_{i}(\mathbf{z})$ and set $\mathfrak{P}(\mathbb{P})=\Phi(\mathbf{z})$. Although such an interpretation describes the observable situation it fails to determine the appropriate couplings when a measurement process is introduced. That is because this interpretation fails to incorporate necessary information about the generation of the informons, in particular that they occur in tuples, not singly. Each round generates a set of informons $\{n^{i}\}$ and thus a set of local Hilbert space contributions, $\{\phi_{n^{i}}(\mathbf{z})\}$.  In order to keep the contributions from individual processes separate it is necessary to resort to a tuple $(n^{1},n^{2},\ldots,n^{j})$ of informons and a corresponding tuple of causal manifold points $\mathbf{m}_{n^{1}},\mathbf{m}_{n^{2}},\ldots,\mathbf{m}_{n^{j}})$ and of local Hilbert space contributions $(\phi_{n^{1}}(\mathbf{z}),\phi_{n^{2}}(\mathbf{z}),\ldots,\phi_{n^{j}}(\mathbf{z}))$. This yields $\mathfrak{P}(\mathbb{P})=\mathfrak{P}(\otimes_{i} \mathbb{P}_{i})=$

$$\{\!((\Phi_{n^{1}}(\mathbf{z}),\Phi_{n^{2}}(\mathbf{z}),\ldots,\Phi_{n^{j}}(\mathbf{z}))|\text{ over all instances of } \mathbb{P}\}\!=$$

$$\mathfrak{P}(\mathbb{P}_{1})\times\mathfrak{P}(\mathbb{P}_{2})\times\cdots\times\mathfrak{P}(\mathbb{P}_{j})$$.

Thus we arrive at a natural relationship between the exclusive product of processes and the associated process covering map. It follows the form of the usual product of wave functions. The indistinguishability of physical entities implies the indistinguishability of their generating processes and so products of processes that are to represent physical entities shall need to follow the usual quantum mechanical formulation rules.

The above holds for free products as well. Indeed a moment's reflection will suggest that

$$\mathfrak{P}(\hat\otimes_{i} \mathbb{P}_{i})=\mathfrak{P}(\mathbb{P}_{1})\times\mathfrak{P}(\mathbb{P}_{2})\times\cdots\times\mathfrak{P}(\mathbb{P}_{j})$$

\noindent so that $\otimes_{i} \mathbb{P}_{i}$ and $\hat\otimes_{i} \mathbb{P}_{i}$ are also $\Psi$-epistemic equivalent. Again the degeneracy in the PCM means that dynamical information is lost when moving over to the wave function, again suggesting that NRQM is incomplete dynamically.

The situation for interactive products is much more complicated. An interactive product, whether free or exclusive, will possess a sequence tree that is a proper subset of the sequence tree for the corresponding independent free or independent exclusive product. In general there may be no simple algebraic representation that generates this sequence tree and which relates the PCM of the interactive product to the PCMs of the individual sub-processes. The PCM for the interactive product must be generated from the sequence tree itself. In NRQM it is assumed that all possible states of systems can be derived from sums and tensor products of states of sub-systems. That is not true in the process setting. 

For example, the sequence tree for the Schr\"odinger cat situation is such that every time the cyanide cannister is opened, the cat dies. Until then, the cat could live or die. Once dead, the cat never returns to life. The combined process of cat and cannister thus will generate sequences where one may have closed cannister and live or dead cat informons, but once a  cannister informon manifests there will only be dead cat informons.
This does not possess a simple algebraic representation in terms of process sums and products.

The PCM shows that the relative strength of a process $\mathbb{P}$ is not an invariant but instead depends upon the particular causal tapestry that has been generated from the initial causal tapestry during an application of the process. In the asymptotic limit $N,r\rightarrow\infty$, the PCM becomes single valued, yielding the Hilbert space interpretation $\Phi^{t_{P}l_{P}}(\mathbf{z})$. For fixed values of $t_{P},l_{P}$, this function is fixed and so one may define the strength of $\mathbb{P}$, $||\mathbb{P}||$ as $||\mathbb{P}||^{2}=\sum_{n_{{i}}\in \mathcal{I}_{\infty}} l^{3}_{P}\Gamma^{*}_{n_{i}}\Gamma_{n_{i}}$. In the case of sinc interpolation one can show that $\sum_{n_{{i}}\in \mathcal{I}_{\infty}} l^{3}_{P}\Gamma^{*}_{n_{i}}\Gamma_{n_{i}}= ||\Phi(\mathbf{z})^{t_{P}l_{P}}||^{2}=
\int_{M}\Phi^{*}(\mathbf{z})\Phi(\mathbf{z})=
\int_{M}\sum_{n_{i}}\sum_{n_{j}}\Phi^{t_{P}l_{P}*}_{n_{i}}(\mathbf{z})\Phi^{t_{P}l_{P}}_{n_{j}}(\mathbf{z})$ \cite{Zayed}. Hence $||\mathbb{P}||^{2}=||\Phi(\mathbf{z})^{t_{P}l_{P}}||^{2}$ and this process strength is well defined. In the event that the process generates a NRQM wave function, this process strength becomes a conserved quantity, an invariant.

\section{Process and Operators}

The study of the relationship between processes and operators on a Hilbert space is unexplored and promising. Only a few remarks will be offered here.

In the discussion above, the PCM was referenced to a given causal tapestry $\mathcal{I}$ which served as an initial condition upon which the process $\mathbb{P}$ was to act. It is certainly possible that the initial condition could be the empty tapestry but in general it will represent the outcome of the actions of previous processes.

Recall that a causal tapestry $\mathcal{I}$ consists of  informons, each of which has the form $[n]<(\mathbf{m}_{n},\phi_{n}(\mathbf{z}),\Gamma_{n},\mathbf{p}_{n})>\{G_{n}\}$. The function $\phi_{n}(\mathbf{z})$ provides a local contribution from $n$ to a global function $\Phi_{\mathcal{I}}(\mathbf{z})$ defined on the causal manifold $\mathcal{M}$ by $\Phi_{\mathcal{I}}(\mathbf{z})=\sum_{n\in \mathcal{I}} \phi_{n}(\mathbf{z})$. This defines a mapping $\mathfrak{I}$ from the space of causal tapestries $\mathbf{I}$ to the Hilbert space $\mathcal{H}(\mathcal{M})$ by

$$\mathfrak{I}(\mathcal{I})=\Phi_{\mathcal{I}}(\mathbf{z})$$.

The PCM was defined as a map $\mathfrak{P}_{\mathcal{I}}:\Pi\rightarrow \mathcal{P}(\mathcal{H}(\mathcal{M}))$. Here the map is referenced to a given causal tapestry $\mathcal{I}$ but clearly $\mathcal{I}$ is free to roam over the entire space of causal tapestries \textbf{I}. More precisely then we should define $\mathfrak{P}$ as a map $\mathfrak{P}:\Pi\times \text{\textbf{I}}\rightarrow \mathcal{P}(\mathcal{H}(\mathcal{M}))$. If we fix some process $\mathbb{P}\in \Pi$ then we can define a tapestry covering map (TCM) $\mathfrak{P}_{\mathbb{P}}:\text{\textbf{I}} \rightarrow \mathcal{P}(\mathcal{H}(\mathcal{M}))$ in the obvious manner.

Define a generalized operator $\mathcal{G}$ on $\mathcal{H}(\mathcal{M})$ as a mapping $\mathcal{G}:\mathcal{H}(\mathcal{M})\rightarrow \mathcal{P}(\mathcal{H}(\mathcal{M}))$ such that $\mathcal{G}(f+g)\subset\mathcal{G}(f)+\mathcal{G}(g)$ and let $\mathfrak{G}(\mathcal{H}(\mathcal{M}))$ denote the set of generalized operators on $\mathcal{H}(\mathcal{M})$.

For a fixed process $\mathbb{P}$, define a generalized operator $\mathfrak{G}_{\mathbb{P}}$ on $\mathcal{H}(\mathcal{M})$ as follows:

For every $f\in \mathcal{H}(\mathcal{M})$, set $\mathfrak{G}_{\mathbb{P}}(f)=\cup_{\mathcal{I}\in \mathfrak{I}^{-1}(f)} \mathfrak{P}_{\mathbb{P}}(\mathcal{I})$

One thus obtains the following diagram

\begin{displaymath}
\begin{array}{ccc}
\mathbf{I} & \stackrel{\mathfrak{P}_{\mathbb{P}}}{\rightarrow} & \mathcal{P}(\mathcal{H}(\mathcal{M})) \\
\mathfrak{I}\downarrow &  & \downarrow e \\
\mathcal{H}(\mathcal{M}) & \stackrel{\rightarrow}{\mathfrak{G}_{\mathbb{P}}} & \mathcal{P}(\mathcal{H}\mathcal{M}) \\
\end{array}
\end{displaymath} 

\noindent where $e$ is a map such that $h\subset e(h)$.

The problem is that one cannot guarantee that if two causal tapestries $\mathcal{I},\mathcal{I}'$ satisfy $\mathfrak{I}(\mathcal{I})=\mathfrak{I}(\mathcal{I}')$ (that is, they generate the same global Hilbert space interpretation), then $\mathfrak{P}_{\mathbb{P}}(\mathcal{I})=\mathfrak{P}_{\mathbb{P}}(\mathcal{I}')$ (that is, the process $\mathbb{P}$ generates the same collection of global Hilbert space interpretations). A process $\mathbb{P}$ is said to be $\Psi$-\emph{faithful} if $\mathfrak{I}(\mathcal{I})=\mathfrak{I}(\mathcal{I}')$ implies that $\mathfrak{P}_{\mathbb{P}}(\mathcal{I})=\mathfrak{P}_{\mathbb{P}}(\mathcal{I}')$ for all $\mathcal{I},\mathcal{I}'$. In the case of a $\Psi$-faithful process the diagram reduces to the simpler form

\begin{displaymath}
\begin{array}{ccc}
\mathbf{I} & \stackrel{\mathfrak{P}_{\mathbb{P}}}{\rightarrow} & \mathcal{P}(\mathcal{H}(\mathcal{M})) \\
\mathfrak{I}\downarrow &  & \downarrow id \\
\mathcal{H}(\mathcal{M}) & \stackrel{\rightarrow}{\mathfrak{G}_{\mathbb{P}}} & \mathcal{P}(\mathcal{H}\mathcal{M}) \\
\end{array}
\end{displaymath} 

\noindent where $id$ is the identity.

In either case one can associate each process $\mathbb{P}$ with a generalized operator $\mathfrak{G}_{\mathbb{P}}$ on $\mathcal{H}(\mathcal{M})$.

If we assume that the process $\mathbb{P}$ involves an effective interpolation strategy, then one can show \cite{Zayed} that in the limit as $N,r\rightarrow \infty$ each PCM becomes effectively a map from $\Pi$ to $\mathcal{H}(\mathcal{M})$ (or equally, each TCM becomes a map from $\mathrm{I}\rightarrow \mathcal{H}(\mathcal{M})$) since the outcome set is a singleton. Consider now the situation in which the asymptotic limit has been taken. This corresponds to restricting attention to only those processes corresponding to the asymptote, so those processes for which $N,r=\aleph_{0}$ at least. As an aside, in combinatorial game theory it is quite possible to work with varying infinite types, particularly the ordinals. If we restrict then to the subset $\Pi_{\infty}$ of such asymptotic processes then one must also restrict the space of causal tapestries to $\mathrm{I}_{\infty}$, the subset of causal tapestries that are generated by processes within $\Pi_{\infty}$. 
Hence for $\mathbb{P}\in\Pi_{\infty}$ and $\mathcal{I}_{\infty}\in \mathrm{I}_{\infty}$,

$$\mathfrak{P}_{\mathcal{I}_{\infty}}(\mathbb{P})=\{\Phi^{t_{P}l_{P}}(\mathbf{z})\},$$ a singleton set. 
 
It follows that the previous diagrams simplify to

\begin{displaymath}
\begin{array}{ccc}
\mathbf{I}_{\infty} & \stackrel{\mathfrak{P}_{\mathbb{P}}}{\rightarrow} & \mathcal{H}(\mathcal{M}) \\
\mathfrak{I}\downarrow &  & \downarrow e \\
\mathcal{H}(\mathcal{M}) & \stackrel{\rightarrow}{\mathfrak{G}_{\mathbb{P}}} & \mathcal{P}(\mathcal{H}\mathcal{M}) \\
\end{array}
\end{displaymath} 

for general processes and for $\Psi$-faithful processes to

\begin{displaymath}
\begin{array}{ccc}
\mathbf{I}_{\infty} & \stackrel{\mathfrak{P}_{\mathbb{P}}}{\rightarrow} & \mathcal{H}(\mathcal{M}) \\
\mathfrak{I}\downarrow &  & \downarrow id \\
\mathcal{H}(\mathcal{M}) & \stackrel{\rightarrow}{\mathfrak{G}_{\mathbb{P}}} & \mathcal{H}(\mathcal{M}) \\
\end{array}
\end{displaymath} 
  
In this situation, the generalized operator $\mathfrak{G}_{\mathbb{P}}$ becomes a standard operator on $\mathcal{H}(\mathcal{M})$. This is the main value for considering $\Psi$-faithful processes.

\section{The Process Approach to Measurement}

Interactions between processes form the fundamental objects of study and this is reflected in the algebraic structure of the process space. An individual process may be active or inactive, and may act singly or in interaction with other processes. Measurement is considered to be a specific type of interaction between a system process and a specialized measurement process.

A Hilbert space will generally possess multiple distinct bases and so given two such bases $\{\phi_{i}(\mathbf{z})\}$ and $\{\sigma_{i}(\mathbf{z})\}$ we may write any element $\Psi(\mathbf{z})$ as $\Psi(\mathbf{z})=\sum_{i}w_{i}\phi_{i}(\mathbf{z})=\sum_{j}v_{j}\sigma_{j}(\mathbf{z})$. In functional analysis one is free to choose any basis at will and represent any vector within that basis. In the ideal world of mathematics this occurs without any physical implications. In the process perspective, ontology is implied. That is, suppose that the sub-process $\mathbb{P}_{i}$ generates $\phi_{i}(\mathbf{z})$ and $\mathbb{S}_{j}$ generates $\sigma_{j}(\mathbf{z})$. Then the superposition $\oplus_{i}w_{i}\mathbb{P}_{i}$ generates $\sum_{i}w_{i}\phi_{i}(\mathbf{z})$ and $\oplus_{j}w_{j}\mathbb{S}_{j}$ generates $\sum_{j}w_{j}\sigma_{j}(\mathbf{z})$. Hence $\oplus_{i}w_{i}\mathbb{P}_{i}$ and $\oplus_{j}w_{j}\mathbb{S}_{j}$ generate the same wave function $\Psi(\mathbf{z})$. However, $\oplus_{i}w_{i}\mathbb{P}_{i}$ and $\oplus_{j}w_{j}\mathbb{S}_{j}$ are \emph{not} the same process. They are sums of distinct sub-processes and will generate distinct informons and possess different sequence trees. Nevertheless they generate the same wave functions via the PCM. They are  $\Psi$-epistemic equivalent processes but they are ontologically distinct.

The treatment of measurement in NRQM is peculiar and gives rise to the measurement problem. Von Neumann \cite{vonNeumann} postulated two distinct dynamical laws for NRQM. The first is a unitary evolution of the wave function via an operator derived from the Hamiltonian and expressed by the Schr\"odinger equation. The second applies exclusively to the setting of measurement. To each measurement apparatus is associated a Hilbert space operator $A$ and each eigenvalue $\lambda$ of $A$ corresponds to a possible measurement outcome. The eigenvectors correspond to states of the system that remain invariant during a measurement. If a wave function can be described as a superposition of eigenstates for a measurement operator $A$, say $\Psi(\mathbf{z})=\sum_{i}w_{i}\Psi_{i}(\mathbf{z})$
then under a measurement the wave function evolves as $A\Psi(\mathbf{z})=\sum_{i}w_{i}\lambda_{i}\Psi_{i}(\mathbf{z})$. This evolution is clearly not unitary. Moreover in reality each system is observed to transition to a single eigenstate, say $\Psi_{i}(\mathbf{z})$, since only a single measurement value $\lambda_{i}$ is ever observed and under repeated subsequent measurements only the value $\lambda_{i}$ is obtained. This leads to the problem of wave function collapse.

For simplicity let us consider a single particle generated by a process $\mathbb{P}$, interacting with a measurement apparatus generated by a process $\mathbb{M}$ and ignore details of experimental error. The measurement process can be further subdivided into local subprocesses $\mathbb{M}_{j}(\mu_{j})$ corresponding to either the different physical components of the apparatus or perhaps to the different measurement responses of the apparatus. The measurement process is understood as taking place in three stages.

\begin{enumerate}
\item In the first stage, prior to the initiation of any interaction between the quantum system and the measurement apparatus, the two processes act freely, so that the combined process can be represented as $\mathbb{P}\otimes \mathbb{M}=\mathbb{P}\otimes (\bigtriangleup _{j}\mathbb{M}_{j}(\mu_{j}))$. The symbol $\bigtriangleup$ denotes either the interactive sum $\boxplus$ or the interactive product $\boxtimes$ depending upon how the processes generating the measurement apparatus states are conjoined and because there may be superselection rules which constrain the generation of measurement apparatus informons.

\item In the second stage, the particle enters into the region of the measurement apparatus. In this stage there is no exchange of physical attributes such as energy, momentum, spin, mass etc. Instead the measurement apparatus establishes a new set of boundary conditions which results in a purely informational effect upon the process generating the particle. This information results in a $\Psi$-preserving transformation in the processes generating the wave function. This occurs because it is precisely those particle subprocesses that generate wave functions that are eigenstates of the operator corresponding to the measurement apparatus that are preserved under an interaction with the apparatus. Only these subprocesses can survive an interaction with the apparatus for a long enough period of time so that a measurement can take place.  Although the subprocesses generating the wave function have changed, the observed wave function remains unchanged. We may now write this new process as $\mathfrak{M}(\mathbb{P})=\oplus_{j} w_{j}\mathbb{P}_{j}(\lambda_{j})$ where $\mathbb{P}_{j}(\lambda_{i})$ represents the process that generates the eigenfunction $\Psi_{j}^{\lambda_{j}}(\mathbf{z})$ corresponding to the eigenvalue $\lambda_{j}$ of the measurement apparatus. Here $\mathfrak{M}(\mathbb{P})$ denotes a mapping on the process space $\Pi$ which carries $\mathbb{P}$ to the $\Psi$-epistemic equivalent process $\mathfrak{M}(\mathbb{P})$ formed from a superposition of eigenprocesses relative to the measurement process $\mathbb{M}$. Note that $\mathfrak{M}(\mathbb{P}_{i}(\lambda_{i}))=\mathbb{P}_{i}(\lambda_{i})$, so that $\mathfrak{M}^{2}=\mathfrak{M}$, so that $\mathfrak{M}$ is a projection.  Although this interaction is purely informational, it nevertheless results in a transition of processes, which we write as:

\begin{center}
$$\mathbb{P}\otimes \mathbb{M}\rightarrow \mathfrak{M}(\mathbb{P})\otimes \mathbb{M}=(\oplus_{i}w_{i}\mathbb{P}_{i}(\lambda_{i}))\otimes \mathbb{M}=$$
\end{center}

$$\oplus_{i}\bigtriangleup_{j} \mathbb{(P}_{i}(\lambda_{i})\otimes \mathbb{M}_{j}(\mu_{j})).$$

\item In the third stage the quantum system enters into the region in which a physical interaction with the measurement apparatus becomes possible. Now each subprocess of $\mathfrak{M}(\mathbb{P})$ can couple only with certain subprocesses of the measurement apparatus (corresponding to nearby measurement values) which may reflect different basins of attraction so that the combined process at the initiation of measurement will be represented more accurately as  $\mathfrak{M}(\mathbb{P})\boxtimes \mathbb{M} = \bigtriangleup_{j}( \oplus_{i\in H(j)}(w_{i}\mathbb{P}_{i}(\lambda_{i}))\boxtimes\mathbb{M}_{j}(\mu_{j})$ where $H(j)$ is the set of system states that can couple to a given measurement apparatus state ($H(j)$ is a singleton in the case of error free measurements). The interactive product $\boxtimes$ represents the interaction between the system process and the measurement apparatus process, which may result at some point in a transition in the process generating the system.  

In the case of no error this becomes $\mathfrak{M}(\mathbb{P})\boxtimes \mathbb{M} = \bigtriangleup_{i}(w_{i}\mathbb{P}_{i}(\lambda_{i})\boxtimes \mathbb{M}_{i}(\lambda_{i}))$.

The actual initiation of interaction becomes a function of the coupling between system and apparatus. That is, with each action of process there will manifest a system informon, say $[n]<\alpha>\{G\}$. Let $\alpha = (\mathbf{m},\phi,\mathbf{p})$. By assumption this informon will be generated by some process of the form $w_{i}\mathbb{P}_{i}(\lambda_{i})$ for some eigenvalue $\lambda_{i}$, so that its Hilbert space interpretation  will be an interpolation contribution to $w_{i}\Psi_{i}^{\lambda_{i}}(\mathbf{z})$, the eigenfunction corresponding to $\lambda_{i}$. Note that $\phi(\mathbf{m})=w_{i}\Psi_{i}^{\lambda_{i}}(\mathbf{m})$. There will also manifest a measurement apparatus informon, say $[m]<\beta>\{H\}$ which is being generated by some subprocess, say $\mathbb{M}_{j}(\mu_{j})$. The likelihood that the system process will couple to the measurement apparatus process is assumed to depend upon the strength $||w_{i}\mathbb{P}_{i}(\lambda_{i})||^{2}$ of the process $w_{i}\mathbb{P}_{i}(\lambda_{i})$ at that informon and so should be proportional to $l^{3}_{P}\Gamma^{*}_{n}\Gamma_{n}=l^{3}_{P}\phi^{*}(\mathbf{m})\phi(\mathbf{m})=l^{3}_{P}w^{*}_{i}(\Psi_{i}^{\lambda_{i}})^{*}(\mathbf{m})w_{i}\Psi_{i}^{\lambda_{i}}(\mathbf{m})$.
The likelihood that it will fail to couple is thus proportional to $p=1-l^{3}_{P}\Gamma^{*}_{n}\Gamma_{n}=p=1-l^{3}_{P}\phi^{*}(\mathbf{m})\phi(\mathbf{m})$. If it does not couple then a new informon is created and the likelihood that it will not couple will again be proportional to $1-x$ for some non-zero $x$. The probability that it will not couple after two acts is proportional to $p(1-x)$ and this will rapidly tend to zero with successive action. Thus almost certainly at some point a coupling will be initiated between the particle and the measurement apparatus if $t_{P}$ is small enough and the strength is great enough.. 
\item Suppose that this current system informon $n$ couples to the current measurement apparatus informon. Then the global process undergoes a transition to an interactive process which is substantially reduced from the original process, namely

\begin{displaymath}
(kw_{i}\mathbb{P}_{i}(\lambda_{i})\boxtimes \mathbb{M}_{k'}(\lambda_{k'}))\oplus \mathbb{M}'
\end{displaymath}

or in the case of no error to

\begin{displaymath}
(kw_{i}\mathbb{P}_{i}(\lambda_{i})\boxtimes \mathbb{M}_{i}(\lambda_{i}))\oplus \mathbb{M}'
\end{displaymath}
 
\noindent where $k=||\mathbb{P}||/||w_{i}\mathbb{P}_{i}||$ and where $\mathbb{M}'$ represents those residual components of the measurement apparatus with which the system does not interact and which do not contribute to the measurement outcome. Note also that this describes a demolition type measurement. In a non-demolition measurement one would expect that the system process does not undergo a change but as a result of the interaction one of the measurement sub-processes, which is initially inactive,  becomes active. There may be other subtleties as well but the version of the measurement interaction presented here provides a sense of the process approach.

Returning to the example, as a result of the interaction with the measurement apparatus the system process has undergone a transition to the sub-process $\mathbb{P}_{i}(\lambda_{i})$. Note that the value $w_{i}$ has been replaced by $kw_{i}$ which ensures that the strength of the new system process matches that of the original process, thereby ensuring that process strength is conserved as expected within NRQM. Since $\mathbb{P}$ is $\Psi$-epistemic equivalent to $\oplus_{i}w_{i}\mathbb{P}_{i}(\lambda_{i})$ it follows that $||\mathbb{P}||^{2}=||\oplus_{i}w_{i}\mathbb{P}_{i}(\lambda_{i})||^{2}$. Thus in the transitions from $\mathbb{P}\rightarrow \oplus_{i}w_{i}\mathbb{P}_{i}(\lambda_{i})\rightarrow kw_{i}\mathbb{P}_{i}$, the process strength is always conserved.

Conservation of process strength makes sense since it should be an attribute of the system being generated, and if the system now concentrates its informons over a lesser range of eigenstates then its strength too should be concentrated over this new range. The system may spread its strength over a greater or a lesser range but it is always the systems strength unless some strength altering interaction takes place.

The frequency of measurement values depends upon the ratios of local strengths and not on process strengthIn NRQM the wave function is generally taken to have norm $1$ so as to preserve its interpretation as a probability distribution. One may single out in $\Pi$ the set of all processes having strength $1$, denoted as ${}_{1}\Pi$. The original space may be reconstructed as $\Pi=\{\sum_{i}w_{i}\mathbb{P}_{i}|w_{i}\in \mathbb{C},\mathbb{P}_{i}\in {}_{1}\Pi\}$.

The system process has undergone a transition which means that the descriptor no longer refers to the original generative process but only to the current process. As a consequence there will no longer be any play to those informons whose descriptors are for the previous process and its subprocesses. The same is true for the measurement apparatus as its generative process is now restricted to  $\mathbb{M}_{i}(\lambda_{i})\oplus \mathbb{M}'$.
 
If the system exist the measurement apparatus it will be generated by the process $kw_{i}\mathbb{P}_{i}(\lambda_{i})$ so if it is observed again by an identical measurement apparatus it will again couple to the same component of the second measurement apparatus and yield the measurement value $\lambda_{i}$ but if it should be observed by a new apparatus then a new coupling will arise as a result of the transition that will be induced in the generating process by the interaction with the new measurement apparatus.
\end{enumerate} 

To understand how probabilities arise within the process framework, consider the effect of a succession of runs of the system process-measurement process interaction. As a result of a single complete run of a process, a single causal tapestry is generated. The strength of the system process determines the likelihood of coupling between the system and the measurement apparatus processes, and the likelihood that a transition to a single eigenprocess will take place. The greater the strength of the process, the greater the strength at its informons and the greater the likelihood that such a transition takes place. If it does not, the original pair of processes continue to act and a new causal tapestry is generated. This pattern will repeat until the transition takes place in a first past the post manner. The process strength thus determines the rate at which the transition takes place, and thus the number of causal tapestries generated by the original pairing prior to the transition. Unlike in standard NRQM, the strength of the process which is related to the norm of the wave function has a definite meaning and effect.

Once the transition takes place, an interactive coupling between some eigenprocess of the system process and the measurement apparatus process begins to generate causal tapestries, ultimately leading to an observable measurement outcome. The particular outcome observed will depend upon which particular informon triggered off the interactive transition. This coupling depends upon the local process strength at the informon in question. 

If a series of repetitions of the experiment are carried out then the frequency of appearance of $\lambda_{i}$ as the measured value will depend upon the relative frequency of couplings to informons generated by $\mathbb{P}_{i}(\lambda_{i})$. This will be obtained by summing strengths over all possible informons generated by $\mathbb{P}_{i}(\lambda_{i})$ and this will be $\sum_{n_{i}\in L_{i}}l^{3}_{P}w^{*}_{i}w_{i}\Gamma^{*}_{n_{i}}\Gamma_{n_{i}}$ where $L_{i}$ refers to the subcollection of informons corresponding to $\mathbb{P}_{i}(\lambda_{i})$. This is just the strength corresponding to $w_{i}\mathbb{P}_{i}(\lambda_{i})$. In the usual NRQM situation, the norm of the wave functions are all taken to be $1$ so that the strength will be given by $w^{*}_{i}w_{i}$. Thus the probability that the measurement outcome is $\lambda_{i}$ will be proportional to $w^{*}_{i}w_{i}$, the usual NRQM result. 

In the process framework, the observation of a measurement value requires an interaction between the system process to be measured and a suitable measurement process. A specific measurement value will be observed only in the event that a coupling between an informon possessing that measurement value (and therefore generated by a sub-process possessing that value as a property) and the measurement process takes place. Such a coupling will depend upon the strength of the process at that informon and since the specific informon cannot be determined this value will need to be determined by summing over all possible suitable informons. That value will be proportional to the strength of the relevant generating process as noted above. In this way there arises, in an emegrent manner, a probability distribution associated with the act of measurement. This probability is contextual since it depends upon the particular measurement process to which the system process is coupled.
The probability thus generated in the process framework is a standard frequency based probability. There is no need to consider observers or beliefs. The wave function has a realist interpretation as a measure of process strength and the probability is derived from rather than intrinsic to the wave function.

The Born rule is generally given as an interpretational postulate for NRQM but now it can be seen as a direct implication of the process interpretation.

\section{Process Approach to the Paradoxes}

It is not possible to do justice to all of the varied paradoxes that have arisen in NRQM within a single brief section. Here I can at best suggest how the process approach can eliminate these from concern. As seen in the previous section, measurement is not accorded a unique status, rather it is merely a particular kind of interaction. There is no special position accorded to the observer in this model. There is a limited form of non-contextuality, in the sense that the informons that are generated by a process $\mathbb{P}$ are attributed a set of properties that are inherited from $\mathbb{P}$. It is not possible, however, for a single process to generate all possible properties. This is due to the fact that processes are closely associated with operators and processes fail, in general, to commute when applied in sequence, just as do operators. Thus in the process framework informons are assigned a limited set of definite properties, and a change in active process will generate informons possessing a different set. These properties are not, however, directly observable. There can only be ascertained by means of an interaction with a measurement apparatus, and thus a measurement process. A measurement process that is compatible, meaning which commutes, with the system process, will leave the system process invariant and so primitive system processes will yield single measurement values and superpositions of such processes will yield a distribution of values. If the measurement process is incompatible with the system process then only a distribution of values will ever be obtained, even if the system process is primitive. Thus the measurement interaction is contextual.

The observer does not generate reality. The observer merely selects a particular measurement apparatus with which the system in question is to interact. The processes involved generate reality, and process exist independent of any observer. Indeed the observer is just another complex process. In the process framework there is no measurement problem.

There is also no wave-particle duality. The duality apparent in both the behaviour and the description of quantum systems created deep conceptual problems for the founders of quantum mechanics. Recall Bohr's comments on this question \cite{Bohr}

\begin{quote}
... how flawed the simple wave-particle description is. Once light [or a material particle] is in an interferometer, we simply cannot think of it as either a wave or a particle. Nor will melding the two descriptions onto some strange hybrid work. All these attempts are inadequate. What is called for is not a composite picture of light, stitched together out of bits taken from various classical theories. Rather we are being asked for a new concept, a new point of view that will be fundamentally different from those developed from the world of classical physics.
\end{quote} 

Bohr and his followers proposed a decidedly anti-realist conception of reality. Whether a quantum system exhibits wave or particle behaviour became dependent upon the choice of an observer or of an experimental setup. This behaviour was no longer an intrinsic aspect of the quantum system but dependent upon the context that the system found itself in. Wheeler's delayed choice experiment poses an even greater conceptual challenge since the choice of whether a quantum system exhibited particle or wave like behaviour could be left until \emph{after} the system entered the experimental apparatus, and thus presumably after the system itself should have made its choice. 

Underlying all of these conundrums is the assumption that a quantum system must be \emph{either} particle-like or wave-like and \emph{not} anything else. Even Bohr believed that. But this is a problem with the manner in which such behaviour is represented mathematically.  Both viewpoints represent idealizations and extremes - particles having no spatio-temporal extension while waves have complete spatio-temporal extension.

Contrary to Bohr's unduly pessimistic view, there is in fact a middle ground given by process theory. Representing an entire function $f$ using an interpolation expansion  of the form 

\begin{displaymath}
f(x)=\sum_{x_{i}}f(x_{i})T_{x_{i}}sinc_{\omega}(x)
\end{displaymath}

\noindent neatly incorporates both discrete features, arising from the discrete nature of the sampling set $\{x_{i}\}$, and continuous features, arising from the coupling to the sinc wavelets and the global summation.

The process model resolves the wave-particle duality problem by incorporating one additional aspect - namely it imposes a dynamic on the creation of these interpolation samples such that they are generated sequentially and not simultaneously. As a consequence, at each step in the generation process there is a single localized expression of the quantum system - a discrete, particle-like entity, but due to the extremely small scale at which these entities, these actual occasions, manifest they are unobservable to the emergent entities that form our observable reality (although they are observable to the processes that generate them) and as a result it is only the global $\mathcal{H}\mathcal{M})$-interpretation that is observable, and that interpretation is an interpolation of a continuous, wave-like entity. The process model eliminates the false dichotomy between particle and wave. In the process model, each informon possesses both particle and wave aspects. The particle aspects are represented by the embedding into the causal manifold, which interprets the informon as being associated with a specific space-time location. The wave aspects are represented by the local $\mathcal{H}(\mathcal{M})$-interpretation, which interprets each informon as being a fuzzy wave like entity whose intensity is highly concentrated around the embedding point. The local $\mathcal{H}(\mathcal{M})$-interpretation serves as a frame element contributing to a global interpolation of a wave function over the space-time surface into which the informons embed. Both the embedding and the $\mathcal{H}(\mathcal{M})$-interpretation are emergent aspects of the quantum system.  
  
Superpositions were dealt with using the exclusive sum $\oplus$ which ensured that if the process $\mathbb{P}=\oplus_{i}w_{i}\mathbb{P}_{i}$ and the $\mathbb{P}_{i}$ are primitive processes, each with a property vector $\mathbf{p}_{i}$, then  informons generated by $\mathbb{P}$ would each be generated by a single subprocess, say $\mathbb{P}_{i}$ and so would be accorded a single property vector $\mathbf{p}_{i}$. Superpositions are always of processes, never of informons. In the case of the two slit experiment one often sees the wave function of the particle written as $\frac{1}{\sqrt{2}}(\Psi_{L}+\Psi_{R})$ where $\Psi_{L}$ corresponds to the particle passing through the left slit, and $\Psi_{R}$ to the particle passing through the right slit. In process terms, let the left process be $\mathbb{P}_{L}$ and the right process $\mathbb{P}_{R}$. In the process framework it is important to note that passage through one slit or another is not an intrinsic property of the particle. Rather, it is a feature of the boundary conditions. The same intrinsic particle state applies whether the particle passes through the left or the right slit. Thus it would be improper to use the exclusive sum to conjoin the corresponding processes. Instead one must use the free sum $\frac{1}{\sqrt{2}}(\mathbb{P}_{L}\hat\oplus\mathbb{P}_{R})$. In this case, it becomes possible for a given informon to receive contributions from both paths or from only a single path depending upon whether or not the boundary conditions permit such information flow. Thus one obtains particle -like behaviour when one slit is open, and wave-like behaviour when both slits are open. If both slits are open initially, and this will be true even if the choice takes place after the particle has entered the apparatus. 

Note that information flows only in a causally local manner from prior informons to nascent informon and information never flows between informons as they are generated by a process. In the construction of a causal tapestry, information therefore flows only from the prior tapestry and never within the current tapestry as it is being generated. Thus there is, at least in principle, no conflict with relativity, no action at a distance. Of course it will be necessary to construct a proper process model of relativistic quantum mechanics and quantum field theory to be sure that this holds up. 

The process model presents an ontology in which the generation of single informons is governed by causally local information. At the process level, however, there is a limited form of nonlocality which arises for two reasons: 1) the fact that processes are generators of space-time but do not possess space-time structure in themselves and 2) the action of process in which successive informons need not be spatio-temporally local to one another, though this does not involve the transfer of information. Non-locality as observed at the measurement level is an emergent non-locality which arises because of the nature of process interactions, especially the interactive coupling, which destroys the statistical independence of the conjoined processes, and because of the measurement situation which itself is a process interaction.

The emergent probability generated by the process model is inherently non-Kolmogorov. This is illustrated by the two slit case, where informons that can receive information from both subprocesses $\mathbb{P}_{L}$ and $\mathbb{P}_{R}$ will contain tokens of the form $\sum_{i}w^{L}_{i}+ \sum_{j}w^{R}_{j}$ so that in calculating the strength one obtains values of the form 

$$\sum_{i}(w^{L}_{i})^{*}w^{L}_{i} +\sum_{j}(w^{R}_{j})^{*}w^{R}_{j}+\sum_{i}\sum_{j}(w^{L}_{i})^{*}w^{R}_{j} +(w^{R}_{j})^{*}w^{L}_{i}$$

\noindent This is clearly non-Kolmogorov \cite{Khrennikov,Sulis2}.

The Schr\"odinger cat problem highlights another inadequacy of NRQM. In this problem a living cat is placed in a sealed shuttered room with a cannister of cyanide which is released after a radioactive decay takes places. The problem arises because in standard NRQM there is only one form of sum and one form of product. If we convert this to process terms, the standard NRQM formulation asserts that the description of the room comprises a process $\mathbb{P}(D)$ for the detector and a process $\mathbb{P}(C)$ for the cat. These two may be further subdivided into the subprocesses $\mathbb{P}(D_{n})$ for the detector in the non-release state, $\mathbb{P}(D_{r})$ for the detector in the release state, $\mathbb{P}(C_{a})$ for the alive cat, and $\mathbb{P}(C_{d})$ for the dead cat. It is then presumed that these can be independently summed and that an independent product links them. That is, it is assumed that one may write

\begin{displaymath}
\Psi = \frac{1}{\sqrt{2}}[(\mathbb{P}(D_{n})\otimes \mathbb{P}(C_{a})) \oplus (\mathbb{P}(D_{r}) \otimes \mathbb{P}(C_{d}))]
\end{displaymath}

The use of the independent sum implies that on any given step it is possible to play either game, up until the game ends (which here means that the observer unseals the room). In such a case the cat would appear to oscillate randomly between a state of being alive and a state of being dead, or as some would have it, in a weird combination of both.

The problem, however, is that it is impossible for the cat to ever effect a transition from the dead state to the alive state. It can remain indefinitely (more or less if the observer doesn't wait too long) in either the alive or dead state, or transition from alive to dead, but never the converse. As a result it is simply impossible to play the free sum because there are selection rules operating which prevent certain transitions in the sequence tree from taking place. Thus the proper sum is the interacting sum. Moreover it is simply not possible to form a state for the cat such as  $\frac{1}{\sqrt{2}}[\mathbb{P}(C_{a}) \oplus \mathbb{P}(C_{d})]$ on account of these transition rules. Thus the only proper description for the combined state is as an interactive sum $\frac{1}{\sqrt{2}}[\mathbb{P}(C_{a}) \boxplus \mathbb{P}(C_{d})]$. The sequence tree allows repeated play of the alive process but once the dead process gets activated the only allowable moves are of the dead process.

The proper description of the process in the room is therefore

\begin{displaymath}
\mathbb{P}(D)\boxtimes \mathbb{P}(C)=\frac{1}{\sqrt{2}}[(\mathbb{P}(D_{n})\otimes \mathbb{P}(C_{a})) \boxplus (\mathbb{P}(D_{r}) \otimes \mathbb{P}(C_{d}))]
\end{displaymath}    

\noindent Note that the interactive product is used in the initial expansion because there is no interaction between the canister prior to decay and the cat, but there is an interaction between the canister after decay and the cat, namely the cat dies, so the canister and cat processes cannot conjoin freely. 

Note that the choice of the play of the process is entirely driven by the process for the canister since as soon as it is played the cat dies. The play of the cannister process allows for a spontaneous transition from the no release process to the decay process, this transition being determined stochastically by the decay rate. The probability of finding a live cat thus depends upon the spontaneous decay rate and will be equal to the probability that no decay occurs during the period of observation. This will equal the likelihood that over the play of the entire process one only plays the no release process for the cannister.

This problem of classicality more generally is a problem worthy of a paper in its own right and so only a few tentative conjectures are offered here to foster further study.

In NRQM it is simply assumed that the linearity of the Schr\"odinger equation implies that any two solutions $\Psi_{1}$ and $\Psi_{2}$ may be summed to give an ontologically realizable state. It is the case that two processes may be $\Psi$-epistemic equivalent, yet inequivalent in many other dynamical ways. The NRQM formalism as it stands does not distinguish between them and so does not necessarily take into account the presence of super-selection rules that might prevent certain operations being performed on them. 

These super-selection rules effectively mean that some combinations might yield the zero process, for example

\begin{displaymath}
\mathbb{P}_{1}\oplus \mathbb{P}_{2}=\mathbb{O}
\end{displaymath}

In the Schr\"odinger cat example, $\mathbb{P}(D_{r})\otimes \mathbb{P}(C_{a})=\mathbb{O}$.

How do these super-selection rules arise in the first place? They could simply be postulated, although that seems to be a bit of a cheat. Let us look more closely at what sums imply.

Within the process framework, the presence of a sum conjoining two processes $\mathbb{P}_{1},\mathbb{P}_{2}$ implies that they act sequentially. Any sequence of processes is possible, but only one process ever acts during a single round. Staying with the exclusive sum, it is also the case that the two processes never act on the same informon. Generally the exclusive sum is used to represent the situation in which one has a single process type, governed by a single strategy type but possibly where there may be different values available to the properties that may be generated. The individual processes in such a sum are meant to represent different instances of the same process type but with possibly different property values being generated. For example a sum of eigenstates is meant to represent a sum of states for the \emph{same} physical system. The conundrum for classicality is that if we apply the same constraint and assume that in the sum we are representing states of the \emph{same} classical system, then we are faced with asserting that the system exists simultaneously in two distinct classical states, something that simply is never observed.

The way out in the discussion of the Schr\"odinger cat problem was to insist that in the classical setting one must use the interactive sum, rather than the independent sum. But why exactly is this necessary? One approach is to assert that this is a scale phenomenon, not manifesting at quantum mechanical scales but manifesting at classical scales, when $\hslash\rightarrow 0$. In the model to be presented in the next section it is shown that this limit is necessary to guarantee NRQM as an idealization of the process model (this is needed in the case that the wave functions are not bounded in energy and momentum). The same argument cannot be used to obtain classicality at the same time. So another mechanism must be in play.

One thing that distinguishes classical from quantum systems is their size. Another feature is their complexity. A classical system consists of a large number (more often vast number) of components which engage in complex interactions with one another.

Each process taking part in a classical superposition is in fact a complex algebraic tangle of primitive processes, some of which will represent different states of single physical systems. If there are $M$ distinct systems comprising the classical system then we may consider a complex process $\mathbb{P}$ to be an element of $\Pi$ formed from the set $\{\{\mathbb{P}^{1}_{i}\},\{\mathbb{P}^{2}_{j}\},\ldots,\{\mathbb{P}^{M}_{k}\}\}$. If we have a second classical process $\mathbb{Q}$ based upon the same component subsystems (another issue for the Schr\"odinger cat example since a live cat is continually renewing its subsystems, something that a dead cat does not) then it will be an algebraic combination based on the set $\{\{\mathbb{Q}^{1}_{i'}\},\{\mathbb{Q}^{2}_{j'}\},\ldots,\{\mathbb{Q}^{M}_{k'}\}\}$

In the realization of the conjoined process $\mathbb{P}\oplus \mathbb{Q}$ there is no issue until following a round in which an informon of $\mathbb{P}$ is generated, there follows a round in which an informon of $\mathbb{Q}$ is generated. The converse situation can be described analogously. The informon of $\mathbb{P}$ will consist of a collection of informons $\{n^{1},n^{2},\ldots, n^{M}\}$ corresponding to each of the component subprocesses.  

When $\mathbb{Q}$ now acts, the information residing in these new informons, being informons of the same component subsystems as governed by $\mathbb{Q}$, may trigger changes in the subproceses that comprise $\mathbb{Q}$, thus inducing a transition to a new classical process $\mathbb{Q}'$, as in the Schr\"odinger cat example where a transition from $\mathbb{P}(C_{a})\rightarrow \mathbb{P}(C_{d})$ can take place.

It is equally possible that the information, although it does not result in a wholesale change of classical process, may preclude possible moves on the part of $\mathbb{Q}$, so that only a subtree of the subsequent sequence tree may actually be implemented. If such a restriction occurs, then it immediately follows that we are no longer operating within the independent sum $\mathbb{P}\oplus\mathbb{Q}$ but instead have transitioned to the interactive sum $\mathbb{P}\boxplus\mathbb{Q}$. 
At some point in the evolution of these processes there may arise a condition under which it is possible that either $\mathbb{P}$ or $\mathbb{Q}$ be be unable to act, and thus rendered inactive, leaving only a single classical process. This may also be the case from the beginning, which would force $\mathbb{P}\oplus\mathbb{Q}=\mathbb{P}\oplus\mathbb{Q}=\mathbb{O}$

The above is clearly sketchy, but it does suggest that the absence of macroscopic superpositions is a expression of the complexity of the conjoining of the individual subprocesses that form a macroscopic or classical object. Indeed, it might be reasonable to \emph{define} a (macroscopic or classical) object to be a complex process for which sums of distinct states are not permitted. In other words, an object would be a collection of complex processes $\{\mathbb{C}_{i}\}$, each generating a distinct state of the object, such that $\mathbb{C}_{i}\oplus\mathbb{C}_{j}=\mathbb{O}$ for all $i,j$.

The process model is discrete and finite, and so it is expected that if it can be generalized to incorporate both relativistic quantum mechanics and quantum field theory, then it will avoid the divergences that plague those theories. That would potentially open the door to also incorporating gravitation, since renormalizability would no longer be a necessary requirement.
 
\section{A Concrete Example: Free Path Integral Strategy}

As with many mathematical structures it is beneficial to work with a concrete representation which possesses the formal mathematical structure but also features that make it amenable to analysis. A physical analogy would be the representations of Lie Algebras and Lie Groups. A particularly useful heuristic representation of process is as a two player, co-operative, combinatorial game \cite{Conway}, based on the forcing games used in mathematical logic to generate models \cite{Hodges}. Combinatorial games are quite distinct from the more familiar economic games. Combinatorial games derive their power and significance from their combinatorial structure. The most common combinatorial games involve the application of various \emph{tokens} to different \emph{}, combinations of which form configurations. Chess, Checkers, Go are familiar examples of combinatorial games. A move alters the tokens associated with a position. Most commonly one considers games involving two players who alternate in making moves and each of which possess a distinct set of tokens and moves. Moves are made non-deterministically, meaning that from any position more than one move is possible and the move is freely chosen without any preassigned probability. It is required that games have a definite end either in the form of some set of \emph{} configurations or by reaching a predetermined limit of individual moves. Usually the last player to move is said to \emph{win} the game.  Games are often described by their game tree which connects a given game configuration to the set of configurations that can be obtained from it by applying a move of either player I or player II. A complete play of the game becomes represented as a path through the game tree beginning with some initial configuration and ending with a final configuration. The game tree is the combinatorial game equivalent of the process sequence tree used previously to construct the process covering map. Combinatorial games come equipped with a remarkable variety of algebraic operations, including many different forms of sum and product \cite{Conway1}. Combinatorial games are intuitive and capture all of the essential algebraic structure of the process space. Combinatorial games are often specified by a set of rules which determine which moves are legal. On each turn, each player is free to select any legal move. It is customary in the study of combinatorial games to assign each player a strategy which consists of a set of additional rules determining their moves. Games are generally analyzed so as to be  independent of strategies. In the process case, different strategies may be thought of as representing different dynamics and so it becomes important to study different strategies as well as different generic processes. The notion of weak-epistemic equivalence (among others) becomes important as a tool for separating out classes of strategies for study. For NRQM one is interested in weak-epistemic equivalent strategies capable of generating global wave functions that satisfy particular Schr\"odinger equations. As an aside one should note that the same approach may be applied to other differential equations, so that the process method has applications outside of quantum mechanics. Which particular strategies become worthy of consideration as plausible physical models will depend upon the specifics of the strategy as far as what it might say about the observable dynamics of the system under consideration. The free strategy described below is not meant to be definitive but rather to provide an in-principle demonstration of the validity of the process approach.

The combinatorial game approach provides a purely heuristic representation of a process which facilitates calculations, and is not meant to be ontological. When applied to implementing a process, the players are not meant to represent real entities but if it aides in intuition one may associate Player I with the system under consideration and Player II with some human observer.

For ease of presentation, consider a single non-relativistic particle with mass $m$, in a single eigenstate of the Hamiltonian, and interacting with a potential $V(\mathbf{z})$, which summarizes the effect of the environment. The particle is modeled as a process while the environment process is ignored (being incorporated into $V$). This simplification allows the value of $V$ associated with an informon $n$ to simply be a property of the informon. Since the setting is non-relativistic one can set $\mathcal{M}=\mathbb{R}^{4}$ and take the causal order to be given by $(t_{1},x_{1},y_{1},z_{1})\prec (t_{2},x_{2},y_{2},z_{2})$ iff $t_{1}<t_{2}$.

Player I propagates information forward to the nascent generation while Player II uses this to construct the new informons. Let $\mathcal{I}_{n}$ be the current generation and $\mathcal{I}_{n+1}$ the nascent generation. Let $\mathcal{I}^{p}_{n}=\cup_{i<n} \mathcal{I}_{i}$. The causal tapestry will be constructed as a sublattice of a uniform lattice. This is a game in which the game board is effectively constructed with each play of the game. Informons may thus be labelled by their site value, which will be of the form $\alpha_{n}=(mt_{P},il_{P},jl_{P},kl_{P})$ where $m$ is fixed, referring to the generation number, and $-\infty\leq i,j,k\leq \infty$. The causal distance between $n\in \mathcal{I}_{m}, n'\in \mathcal{I}_{m+1}$ is given by the Euclidean metric applied to the lattice site values. Hence if $\alpha_{n}=(mt_{P},il_{P},jl_{P},kl_{P})$ and $\alpha_{n'}=((m+1)t_{P},i'l_{P},j'l_{P},k'l_{P})$ then $d(n,n')^{2}=d(\alpha_{n},\alpha_{n'})= t_{P}^{2}+((i-i')^{2}+(j-j')^{2}+(k-k')^{2})l_{P}^{2}$. Note that one could easily set $\mathcal{M}=\mathbb{M}^{4}$, the 4-dimensional Minkowski space with causal order given via the Minkowski metric to obtain the relativistic case.

The process strength $\Gamma_{n}$ of an informon $n$ is thought of as propagating forward to subsequent informons as a discrete (possibly dissipative) wave. The contribution to the next informon $n'$ will depend upon the causal distance between them. If the Lagrangian for the particle is $\mathcal{L}=\frac{p^{2}}{2m}+V$ then each contribution will take the form $e^{(i/\hbar)\{ \frac{md(n,n')^{2}}{2t_{P}^{2}}+V(n)\}t_{P}}$ or alternatively, $e^{(i/\hbar)\{\frac{md(\alpha_{n},\alpha_{n'})^{2}}{2t_{P}^{2}}+V(n)\}t_{P}}$. Let $\mathcal{L}(n,n')=\frac{md(n,n')^{2}}{2t_{P}^{2}}+V(n)$. 
Note that this is derived from information residing solely within the causal tapestry and does not depend upon the causal manifold or $\mathcal{H}(\mathcal{M})$-interpretations. This is in keeping with the view that the physics  takes place on the causal tapestry and the interpretations reflect the point of view of the human observer or of some theory and are not real in themselves but simply constitute idealized descriptions.

The $\mathcal{H}(\mathcal{M})$-interpretation is constructed by means of sinc interpolation, chosen because of its simplicity, its effectiveness for a large class of $L^{2}$ functions on $\mathcal{H}(M)$, and for the richness of the available literature dealing with the range of functions that can be interpolated, convergence properties, error properties and so on. Sinc interpolation requires the use of a lattice embedding into $\mathcal{M}$, which admittedly is unrealistic, but Maymon and Oppenheim \cite{Maymon} have shown that even if the actual embedding point is off lattice, for small errors this will still provide a highly accurate approximation of the actual function, and thus in the situations to be considered in the free path integral strategy, this is a \textquoteleft good enough\textquoteright\; assumption. A more realistic model would require the use of non-uniform embeddings and more sophisticated interpolation techniques, such as Fechtinger-Gr\"ochenik theory \cite{Zayed}, but the details would unnecessarily complicate the presentation. The important point is that the physics is independent of the interpolation scheme which merely serves to translate it into a standard quantum mechanical form, and thus it is somewhat arbitrary and dependent upon the particular circumstances. Sinc interpolation is good enough for the in-principle demonstration presented here. 

The class of functions that can be interpolated by sinc interpolation depends upon the values of $t_{P}$ and $l_{P}$ and will lie within the class $B_{\sigma}^{n}$ (these are the so-called band-limited functions, meaning that their Fourier transforms have support within the bounded region $[-\sigma,\sigma]$). This class is smaller than $L^{2}(\mathcal{H}(M))$ but in general this is not a problem because in reality all quantum systems possess bounded energy and momentum which determine the value of $\sigma$ through $E=\hslash\sigma=2\hslash\pi/t_{P}$ and $p=\hslash\sigma=2\hslash\pi/l_{P}$. Fixing $\sigma$ to be finite is equivalent to asserting the existence of an ultraviolet cutoff, a reasonable assumption when considering the behaviour of any physically real particles. Thus one advantage of sinc interpolation is that a natural ultraviolet cutoff exists by virtue of the nature of the interpolation process and does not need to be specified as an ad hoc assumption. Since quantum mechanics deals mostly with $L^{2}$ functions we need consider only the classes $B^{2}_{\sigma}$ and I will denote $L^{2}(\mathcal{H}(M))$ restricted to $B^{2}_{\sigma}$ by $\mathcal{H}_{\sigma}(\mathcal{M})$.

A basic strategy which provides an in-principle demonstration of the power of the method is that of the Bounded Radiative Uniform Sinc Path Integral Strategy $(\mathfrak{P}\mathfrak{I}$). The path integral strategy is specified by the parameters $R,r,N,t_{P},l_{P}$ and by 

\begin{enumerate}
\item $\Delta$ (distance bound): arbitrary, determines maximum causal distance of information transmission.
\item $\rho$ (approximation measure): arbitrary, set by the observer according to mathematical or experimental considerations.
\item  $\delta$ (approximation accuracy): arbitrary but bounded by experimental measurements
\item $\omega$ (band limit frequency): bounded by upper limits of energy and momentum of the quantum system
\item $\mathcal{L}$ (Lagrangian): determined by the particulars of the quantum system
\item $p$ (set of properties): here energy, momentum
\end{enumerate}

The informons of the causal tapestry $\mathcal{I}_{n}$ will be embedded into a sub-lattice of the space-like hyper-surface (or time slice) $\{nt_{P}\}\times \mathbb{R}^{3}$ in $\mathcal{M}$. The embedding lattice in $\mathcal{M}$ will thus take the general form $(nt_{P},il_{P},jl_{P},kl_{P})$ for integers $n,i,j,k$. The embedding point in $\mathcal{M}$ of an informon $n$ will be denoted $\mathbf{m}_{n}$. In the simplified version presented here, $\alpha_{n}=\mathbf{m}_{n}$ but this is not true generally. In fact it is only necessary that the embedding into $\mathcal{M}$ preserve the causal order and that  the causal distance lies within certain tolerance limits from that on the causal tapestry. In the relativistic case $\mathcal{I}_{n}$ will embed into the causal manifold $\mathcal{M}$ as a sub-lattice of some space-like hyper-surface and the embedding may or may not preserve causal distances (the error being bound by some parameter $\epsilon$.

The path integral strategy for a single short round proceeds as follows:

\begin{enumerate}
\item Player I moves first. Player I non-deterministically chooses any informon $[n]<\alpha_{n}>\{G_{n}\}=[n]<(\mathbf{m}_{n},\phi_{n},\alpha_{n},S_{n},V(n))>\{G_{n}\}$ from the current tapestry $\mathcal{I}_{m}$ which has not previously been played in this round, where $\mathbf{m}_{n}$ is the $\mathcal{M}$ embedding, $\phi_{n}(\mathbf{z})$ is the local Hilbert space contribution, $\alpha_{n}$ is a lattice site, $\Gamma_{n}$ is the strength of the generating process at $n$ and $V(n)$ is the value of the potential at $n$ .

\item If there is an informon $[n']<\alpha_{n'}>\{G_{n'}\}$ currently in play in the new tapestry $\mathcal{I}_{m+1}$ then Player II tests whether $d(n,n')<\Delta$ in the new tapestry $\mathcal{I}'$. If the bound is exceeded, play reverts back to step I, otherwise it proceeds. If there is no current informon then Player II chooses a label $n'$ not previously used and selects a lattice site $((m+1)t_{P},i'l_{P},j'l_{P},k'l_{P})$ not previously used such that $d(\alpha_{n},\alpha_{n'})<\Delta$ and creates a new informon $[n']<\alpha_{n'}>\{\}$. 
\item Player I next updates the content set. If the new inform already possesses a content set $G_{n'}$, then Player I replaces $G_{n'}$ with $G_{n'} \cup \hat G_{n} \cup \{[n]<\alpha_{n}>\{G_{n}\}\}$ ($\hat G_{n}$ is an order theoretic up-set of $G_{n}$) and checks to ensure that all necessary order conditions are satisfied. If the new informon is nascent, then Player I simply sets $G_{n'} = \hat G_{n} \cup \{[n]<\alpha_{n}>\{G_{n}\}\}$. The content set determines what prior information is permitted in constructing tokens. It only includes informons from the past causal cone of the informon. In the case of NRQM, it turns out that only informons from the current tapestry are needed since the relevant information is already incorporated into their $\mathcal{H}(M)$-interpretations. Thus it suffices if $G_{n'}$ or $\emptyset$ is replaced with $G_{n'} \cup \{[n]<\alpha_{n}>\{G_{n}\}\}$ or $\{[n]<\alpha_{n}>\{G_{n}\}\}$ respectively. Note that in this case the causal consistency criteria are trivially satisfied.

\item  Player II next determines the causal manifold embedding. If the nascent informon $n'$ already possesses a causal manifold embedding, then Player II does nothing. Otherwise Player II sets $\mathbf{m}_{n'}=\alpha_{n'}$ and the nascent informon becomes $[n']<\mathbf{m}_{n'},\alpha_{n'}>\{G_{n'}\}$.

\item Player I next constructs a token representing the information passing from $n$ to $n'$ and to be used to form the local Hilbert space contribution at $n'$. Denote this token as $\mathcal{T}_{n'n}$. Let $\tilde S[n',n]= (\frac{md(n',n)^{2}}{2t_{P}^{2}}+V(n))t_{P}$. Let $T_{n}$ denote the set of tokens on $n$. Let $\phi_{n}$ denote the sum of the tokens on $n$, that is $\phi_{n}=\sum\{\mathcal{T}_{nm}|\mathcal{T}_{nm}\in T_{n}\}$. In what follows $\Phi_{n}(\mathbf{z})$ will refer to the local $\mathcal{H}(\mathcal{M})$-interpretation of $n$.  The relationship between these two is $\phi_{n}=(1/A^{3})\Phi_{n}(\mathbf{m}_{n})$, where $A$ is the path integral normalization factor described by Feynman and Hibbs \cite{Feynman} which is appropriate to the current Lagrangian and initial and boundary conditions. The reason for this will become apparent later.  Define the propagator $P_{n'n}=(l_{P}^{3}/A^{3})e^{i\tilde S[\alpha_{n'},\alpha_{n}]/\hbar}$. Then Player I places a token $\mathcal{T}_{n'n}=P_{n'n}\phi_{n}$ on the site $\alpha_{n}$. If there already is a set $T_{n'}$ of tokens on informon $n'$ then replace it by $T_{n'}\cup\{T_{n'n}\}$.

\item Finally Player II must determine the $\mathcal{H}(\mathcal{M})$-interpretation. If $\mathbf{z}=(t,x,y,z)$ and $\mathbf{m}_{n'} = ((nt_{P},ml_{P},rl_{P},sl_{P}))$ define $T_{\mathbf{m}_{n'}}sinc_{t_{P},l_{P}}(\mathbf{z})=$

\begin{displaymath}
sinc\left(\frac{\pi (t-nt_{P})}{t_{P}}\right) sinc\left(\frac{\pi (x-ml_{P})}{l_{P}}\right)
\end{displaymath}
\begin{displaymath}
\times sinc\left(\frac{\pi (y-rl_{P})}{l_{P}}\right)sinc\left(\frac{\pi (z-sl_{P})}{l_{P}}\right)
\end{displaymath}

Player II constructs the $\mathcal{H}(\mathcal{M})$-interpretation by coupling the tokens on the site to a suitable interpolation function, which in the current strategy utilizes a sinc function given as $A^{3}T_{\mathbf{m}_{n}}sinc_{t_{P}l_{P}}(\mathbf{z})$. If the new informon has just been formed, then the $\mathcal{H}(\mathcal{M})$-interpretation is given as $\Phi_{n'}(\mathbf{z}) = \mathcal{T}_{n'n}A^{3}T_{\mathbf{m}_{n'}}sinc_{t_{P},l_{P}}(\mathbf{z})$. If the informon already possesses a $\mathcal{H}(\mathcal{M})$-interpretation, $\Phi_{n'}(\mathbf{z})$, then replace it by the new $\mathcal{H}(\mathcal{M})$-interpretation $\Phi_{n'}(\mathbf{z})+\mathcal{T}_{n'n}A^{3}T_{\mathbf{m}_{n'}}sinc_{t_{P},l_{P}}(\mathbf{z}).$

In other words, add the new token to the collection, sum the token values and couple the sum to the interpolation wavelet. 

\item If no further tokens can be added (either no other contributing sites exist or an external limit has been reached), the round ends and a new round begins.
\end{enumerate}

Play continues until the allotted number of allowed game steps has been reached. At the end of play a new causal tapestry $\mathcal{I}_{m+1}'$ has been created and the old causal tapestry $\mathcal{I}_{m}$ is eliminated, formally becoming a part of $\mathcal{I}^{p}_{m+1}$, the collection of prior tapestries. Any relevant information from $\mathcal{I}_{m}$ now resides within the content sets of the informons of $\mathcal{I}_{m+1}$. Let $n'$ denote an informon of $\mathcal{I}_{m+1}$. Let $L_{n'}$ denote the set of all informons from $\mathcal{I}_{n}$ that contribute tokens to the formation of $n'$. Equally, $L_{n'}$ is the set of all informons from $\mathcal{I}_{n}$ that form vertices in $G_{n'}$.  The local $\mathcal{H}(\mathcal{M})$-interpretation of $n'$ may now be written as 
$\Phi_{n'}(\mathbf{z})= \sum_{n\in L_{n'}}\mathcal{T}_{n'n}A^{3}T_{\mathbf{m}_{n'}}sinc_{t_{P},l_{P}}(\mathbf{z})$

The global $\mathcal{H}(\mathcal{M})$-interpretation on $\mathcal{M}$ is formed by summing the local contributions over all of $\mathcal{I}_{m+1}$, that is $\Phi^{m+1}(\mathbf{z})=\sum_{n'\in \mathcal{I}_{m+1}}\Phi_{n'}(\mathbf{z})$. One may restrict this to the $t=(m+1)t_{P}$ hyper-surface, obtaining, as will be shown below, a highly accurate approximation to the standard quantum mechanical wave function on the hyper-surface. Note that fixing $t=m+1$ causes the time based sinc term to take the value 1 and one indeed obtains a function on the hyper-surface. This approximation will be less accurate when extended to the entirety of $\mathcal{M}$. To achieve greater accuracy requires either summing over the content sets of $\mathcal{I}_{m+1}$, i.e. $\Phi^{m+1,c}_{n'}(\mathbf{z})= \sum_{n\in G_{n'}, n'\in I_{n'}}\Phi_{n}(\mathbf{z})$ or over all of $\mathcal{I}_{m+1}\cup \mathcal{I}_{p}$, $\Phi^{m+1,p}_{n'}(\mathbf{z})= \sum_{n\in \mathcal{I}_{m+1}\cup I_{p}}\Phi_{n}(\mathbf{z})$.

A superposition of eigenstates may be generated by an exclusive sum of primitive processes, each corresponding to a single eigenstate. The global wave function may be obtained by interleaving the informons generated by the sub-processes, either on the same lattice, in which case a slightly different interpolation function must be used (or the missing values determined using the scheme of \cite{Marks}) or the informons of distinct processes may be placed on distinct lattices.

\section{Formal Proof of Emergent NRQM}

In the previous section the assertion was made that NRQM can be viewed as an effective theory arising in the asymptotic limit as $N,r\rightarrow \infty$ and $t_{P},l_{P}\rightarrow 0$. To prove this consider the following.

Assume that the particle is generated by a primitive process in an eigenstate of its Hamiltonian. Let $\mathcal{I}_{0}$ denote the initial generation for the particle process $\mathbb{P}$ and assume that on this generation the process strengths $\Gamma_{n}$ correspond to the values of the wave function sampled at the embedding points, i.e $\Gamma_{n}=\Psi(\mathbf{m}_{n})$. 

Parzen's theorem states that if $f(t_{1},\ldots, t_{N})$ is a function band limited to the $N$-dimensional rectangle $B=\prod_{i=1}^{N} (-\sigma_{i},\sigma_{i})$, $\sigma_{i}>0$, $i=1,\ldots,N$ so that its Fourier transform $F(\omega_{1},\ldots,\omega_{N})$ is such that

\begin{displaymath}
\int_{-\sigma_{1}}^{\sigma_{1}}\cdots \int_{-\sigma_{N}}^{\sigma_{N}}|F(\omega_{1},\ldots,\omega_{N})|^{2}d\omega_{1}\cdots d\omega_{N}<\infty,
\end{displaymath}

\noindent $F(\omega_{1},\ldots,\omega_{N})=0$ for $|\omega_{x}|>\sigma_{k}$, $k=1,\ldots, N$, and $\pi k_{i}/\sigma_{i}=\hat k_{i}$, then $f(t_{1} ,\ldots…,t_{N} )=$

\begin{displaymath}
\sum^{\infty}_{k_{1} =-\infty}\negthickspace\cdots\negthickspace\negthickspace\negthickspace \sum^{\infty}_{k_{N}=-\infty} \negthickspace\negthickspace f\left(\hat k_{1} ,.,\hat k_{N}\right)s( \sigma_{1}(t_{1}- –\hat k_{1}) ).. s(\sigma_{N}(t_{N}-\hat k_{N} ))
\end{displaymath}

Therefore by Parzen's theorem, on $\mathcal{I}_{0}$, $\Phi^{0}(\mathbf{z})=$

$$\sum_{n\in \mathcal{I}_{0}}\Gamma_{n}A^{3}T_{\mathbf{m}_{n'}}sinc_{t_{P},l_{P}}(\mathbf{z})=$$

$$\sum_{n\in \mathcal{I}_{0}}\Psi(\mathbf{m}_{n})A^{3}T_{\mathbf{m}_{n'}}sinc_{t_{P},l_{P}}(\mathbf{z})\approx \Psi(\mathbf{z})$$.

Assume that the process has generated all generations up to and including $m+1$. Let $g_{n'}(\mathbf{z})=A^{3}T_{\mathbf{m}_{n'}}sinc_{t_{P},l_{P}}(\mathbf{z})$. Recall that $\Phi^{m+1}(\mathbf{s})=\sum_{n\in \mathcal{I}_{m+1}}\Phi_{n}(\mathbf{z})$.

Hence one can write 

$$\Phi^{m+1}(\mathbf{z})=\sum_{n'\in \mathcal{I}_{m+1}}\sum_{n\in L_{n'}}\mathcal{T}_{n'n}g_{_{n^{m+1}}}(\mathbf{z})$$ 

If we assume the convention that $\mathcal{T}_{n'n}=0$ if $n$ does not propagate information to $n'$ then we can rewrite the above as 

$$\Phi^{m+1}(\mathbf{z})=\sum_{n^{m+1+}\in \mathcal{I}_{m+1}}\sum_{n^{m}\in \mathcal{I}_{m}}\mathcal{T}_{n^{m+1}n^{m}}g_{n^{m+1}}(\mathbf{z})$$ 
$$=\sum_{n^{m+1}\in \mathcal{I}_{m+1}}\sum_{n^{m}\in \mathcal{I}_{m}}\mathcal{P}_{n^{m+1}n^{m}}\phi_{n^{m}}g_{n^{m+1}}(\mathbf{z})$$

$$=\sum_{n^{m+1}\in \mathcal{I}_{m+1}}\sum_{n^{m}\in \mathcal{I}_{m}}\mathcal{P}_{n^{m+1}n^{m}}\negthickspace\negthickspace\negthickspace\sum_{n^{m-1}\in \mathcal{I}_{m-1}}\negthickspace\negthickspace\negthickspace\mathcal{T}_{n^{m}n^{m-1}}g_{n^{m+1}}(\mathbf{z})=$$

$$\negthickspace\sum_{n^{m+1}\in \mathcal{I}_{m+1}}\sum_{n^{m}\in \mathcal{I}_{m}}\mathcal{P}_{n^{m+1}n^{m}}\negthickspace\negthickspace\negthickspace\negthickspace\sum_{n^{m-1}\in \mathcal{I}_{m-1}}\negthickspace\negthickspace\negthickspace\negthickspace\mathcal{P}_{n^{m}n^{m-1}}\phi_{n^{m-1}}g_{n^{m+1}}(\mathbf{z})$$

Continuing one arrives at

\begin{widetext}
$$\Phi_{m+1}(\mathbf{z}) = \sum_{n^{m+1}\in \mathcal{I}_{m+1}}\cdots \sum_{n^{0}\in \mathcal{I}_{0}}\mathcal{P}_{n^{m+1}n^{m}}\cdots\mathcal{P}_{n^{1}n^{0}}\phi_{n{0}}g_{{n^{m+1}}}(\mathbf{z})=$$
$$\sum_{n^{m+1}\in \mathcal{I}_{m+1}}\cdots \sum_{n^{0}\in \mathcal{I}_{0}}\mathcal{P}_{n^{m+1}n^{m}}\cdots\mathcal{P}_{n^{1}n^{0}}\Psi(\mathbf{m}_{n^{0}})g_{n^{m+1}}(\mathbf{z})=$$
$$\sum_{n^{m+1}\in \mathcal{I}_{m+1}}\sum_{n^{m}\in \mathcal{I}_{m}}\cdots \sum_{n^{0}\in \mathcal{I}_{0}}\frac{l_{P}^{3}}{A^{3}}e^{\frac{i}{\hbar}\tilde S[\alpha_{n^{m+1}},\alpha_{n^{m}}]}\frac{l_{P}^{3}}{A^{3}}e^{\frac{i}{\hbar}\tilde S[\alpha_{n^{m}},\alpha_{n^{m-1}}]}\times \cdots\times$$
$$\frac{l_{P}^{3}}{A^{3}}e^{\frac{i}{\hbar}\tilde S[\alpha_{n^{1}},\alpha_{n^{0}}]}\Psi(\mathbf{m}_{n^{0}})g_{n^{m+1}}(\mathbf{z})=$$
$$\sum_{n^{m+1}\in \mathcal{I}_{m+1}}\sum_{n^{m}\in \mathcal{I}_{m}}\cdots \sum_{n^{0}\in \mathcal{I}_{0}}e^{\frac{i}{\hbar}\tilde S[n^{m+1},n^{m}]+\tilde S[n^{m},n^{m-1}]+\cdots +\tilde S[n^{1},n^{0}]}\times $$
$$\overbrace{\frac{l_{P}^{3}}{A^{3}}\frac{l_{P}^{3}}{A^{3}}\cdots \frac{l_{P}^{3}}{A^{3}}}^{m+1}\Psi(\mathbf{m}_{n^{0}})g_{n^{m+1}}(\mathbf{z})=$$
$$\sum_{n^{m+1}\in \mathcal{I}_{m+1}}\sum_{n^{m}\in \mathcal{I}_{m}}\cdots \sum_{n^{0}\in \mathcal{I}_{0}}e^{\frac{i}{\hbar}\mathcal{L}[n^{m+1},n^{m}]t_{P}+ \mathcal{L}[n^{m},n^{m-1}]t_{P}+\cdots + \mathcal{L}[n^{1},n^{0}]t_{P}}\times \overbrace{\frac{l_{P}^{3}}{A^{3}}\frac{l_{P}^{3}}{A^{3}}\cdots \frac{l_{P}^{3}}{A^{3}}}^{m+1}\Psi(\mathbf{m}_{n^{0}})g_{n^{m+1}}(\mathbf{z})\approx$$
$$\sum_{n^{m+1}\in \mathcal{I}_{m+1}}\sum_{n^{m}\in \mathcal{I}_{m}}\cdots \sum_{n^{0}\in \mathcal{I}_{0}}e^{\frac{i}{\hbar}S[\alpha_{n^{m+1}},\alpha_{n^{0}}]}
\overbrace{\frac{l_{P}^{3}}{A^{3}}\frac{l_{P}^{3}}{A^{3}}\cdots \frac{l_{P}^{3}}{A^{3}}}^{m+1}\Psi(\mathbf{m}_{n^{0}})g_{n^{m+1}}(\mathbf{z})\approx$$
$$\sum_{n^{m+1}\in \mathcal{I}_{m+1}}\sum_{n^{0}\in \mathcal{I}_{0}}\int_{I_{i-1}}\cdots \int_{I_{1}}e^{\frac{i}{\hbar}S[\alpha_{n^{m+1}},\alpha_{n^{0}}]}\overbrace{\frac{d\alpha_{n^{m+1}}}{A^{3}}\cdots \frac{d\alpha_{n^{1}}}{A^{3}}}^{m}\frac{l_{P}^{3}}{A^{3}}\Psi(\mathbf{m}_{n^{0}})g_{n^{m+1}}(\mathbf{z})$$
\end{widetext}

\noindent where the $I_{i}$ refers to the continuous extension of the sub-lattice upon which $\mathcal{I}_{i}$ is defined, i.e. $I_{i}=\{it_{P}\}\times \mathbb{R}^{3}$, the action integral has been taken over the piecewise linear path $\alpha_{n^{0}},\alpha_{n^{1}},\ldots,\alpha_{n^{m+1}}$ on the continuous lattice extension $[0,(m+1)t_{P}]\times \mathbb{R}^{3}$ and where the final step is obtained approximating each discrete sum by an integral.

Now as $N,r\rightarrow \infty$, the number of informons from $\mathcal{I}_{k}$ contributing to any informon of $\mathcal{I}_{k+1}$ grows to infinity for each $k$ and a moment's reflection will suggest therefore that in the limit every possible path between the informons of $\mathcal{I}_{0}$ and the informons of $\mathcal{I}_{m+1}$ will be included in the calculation. The entirety of each causal tapestry will be connected to all of the other causal tapestries. As $t_{P}\rightarrow 0$, the temporal spacing between lattice slices decreases, so not only does the total number of lattice slices increase, but the number of lattice slices between $\alpha_{0}$ and $\alpha_{i}$ increases while their distance decreases. Under such circumstances, according to Feynman and Hibbs \cite{Feynman} the product of the integrals above converges to the path integral between the points $\alpha_{0}$ and $\alpha_{i}$. Substituting this into the above one obtains $\Phi^{m+1}(\mathbf{z})\approx$

\begin{widetext}
\begin{multline*}
 \sum_{n^{m+1}\in \mathcal{I}_{m+1}}\sum_{n^{0}\in \mathcal{I}_{0}}K(\alpha_{n^{m+1}},\alpha_{n^{0}})\Psi(\mathbf{m}_{n^{0}})\frac{l_{P}^{3}}{A^{3}}g_{n^{m+1}}(\mathbf{z})=
\negthickspace\sum_{n^{m+1}\in \mathcal{I}_{m+1}}\sum_{n^{0}\in \mathcal{I}_{0}}K(\alpha_{n^{m+1}},\alpha_{n^{0}})\Psi(\mathbf{m}_{n^{0}})l_{P}^{3}T_{\mathbf{m}_{n^{m+1}}}sinc_{t_{P},l_{P}}(\mathbf{z})\\
\approx \sum_{n^{m+1}\in \mathcal{I}_{m+1}}\int_{I_{0}}K(\alpha_{n^{m+1}},\alpha_{n^{0}})\Psi(\alpha_{0})d\alpha_{0}T_{\alpha_{i}}sinc_{t_{P}l_{P}}(\mathbf{z})=\\
\sum_{n^{m+1}\in \mathcal{I}_{m+1}}\Psi(\alpha_{n^{m+1}})T_{\alpha_{i}}sinc_{t_{P}l_{P}}(\mathbf{z})=\Psi(\mathbf{z})
\end{multline*}
\end{widetext}

\noindent where the final step again uses Parzen's theorem. Technically, the final equality holds on the $t=(m+1)t_{P}$ time slice and the relation holds approximately on that portion of $\mathcal{M}$ for which $t\leq (m+1)t_{P}$ but only weakly on the forward portion for which $t>(m+1)t_{P}$.

\section{Errors}

The next question to address is the goodness of fit between the wave function as determined by this model and the wave function as calculated using the usual path integral or Schr\"odinger equation methods. Goodness of fit is a more accurate term because the truly important question is not whether it accurately matches the NRQM wave function but rather how well it satisfies the Schr\"odinger equation and provides the essential statistical relations. Comparison to the NRQM wave function is made by using the discrepancy measure $\rho$ or by substituting into the appropriate Schr\"odinger equation and examining for goodness of fit. This in turn determines whether or not the game is a win for Player II. If it is, then we can say that the reality game  $\mathfrak{R}(N,r,\rho,\delta,t_{P},l_{P},\omega,L,\Sigma,p)$ generates the wave function to accuracy $\delta$.

The discrepancy between the global $\mathcal{H}(\mathcal{M})$-interpretation given above and the standard NRQM wave function depends upon the accuracy of the approximation to the integral $\int_{\mathcal{M}_{t}} K(\mathbf{x}_{j'},\mathbf{x}_{j})\phi_{j}(\mathbf{x}_{j}) d\mathbf{x}_{j}$, the deviations from uniformity of the causal embedding points, the number $r$ of current informons contributing information to any nascent informon as well as the values of $t_{P},l_{P}$.  This is a difficult problem to assess in general but results are available in special cases. For example, in one dimension, if the wave function $\Psi$ satisfies $|\Psi|\leq M|t|^{-\gamma}$ for $0<\gamma\leq 1$, $|\int_{\mathcal{M}_{t}} K(x_{j'},x_{j})\phi_{j}(x_{j}) dx_{j}-\Psi(x_{j'})|\leq \epsilon$, the discrepancy between each embedding point and its ideal lattice embedding point is less that $\delta$, and the truncation number $r=2[W^{1+1/\gamma}+1]+1$, then according to a theorem of Butzer (see Zayed), the error $E$ satisfies

\begin{displaymath}
||E||_{\infty}\leq -K(\Psi,\gamma,\epsilon/l_{P},\delta/l_{P})l_{P}\ln l_{P}
\end{displaymath}

\noindent where
\begin{widetext}
$K=(1+\frac{1}{\gamma})\left\{\sqrt{5}e\left[ (\frac{14}{\pi}+\delta/l_{P}+\frac{7}{3\sqrt{5}\pi})||\Psi^{(1)}||_{\infty}+\epsilon/l_{P}\right]+\\
 6e(M+||\Psi||_{\infty}) \right\}$
\end{widetext}

Hence, $||E||_{\infty}\approx 10^{-33}K$ if $l_{P}$ is the Planck length.

In the ideal case in which the kernel sum equals the kernel integral, and the causal embedding is uniform one can use the Yao and Thomas theorem \cite{Zayed}. to find that the error roughly $T$ satisfies

\begin{displaymath}
|T(\mathbf{z})|\leq \frac{8\max_{\mathcal{M}'} |\Psi(x,y,z)|}{(2\pi)^{3}(\Delta/l_{P})^{3}}\approx 10^{-98}/\Delta \;(m^{-3})
\end{displaymath}

if $l_{P}$ is the Planck length.
 
The appropriate strategies and interpolation methods are a matter for future study and comparison to empirical data. 

\section{Conclusion}

The process framework offers an ontological model for NRQM which is not simply a re-interpretation. The process algebra within which the model is defined possesses a richer algebraic structure than does the standard Hilbert space. It possesses 8 distinct forms of sum and product plus a concatenation operation. The connection to the Hilbert space is through the PCM, which is a covering or quotient map, and thus essential information is lost in making the transition. The process model is a generative or emergent model and contains dynamical information that is not present in the Hilbert space. NRQM may be statistically complete but it is not dynamically complete, and the process model fills in some of the gaps. 

The process model provides a self contained and consistent, realist ontology which is generative, discrete, finite, and intuitive. It appears capable of resolving many, if not all, of the paradoxes that plague quantum mechanics. If it can be extended to relativistic quantum mechanics and quantum field then it will avoid the divergences, obviate the need for renormalizability and thus potentially open the door to a unification of quantum mechanics and gravitation. 

It reduces the degree of non-locality at the lowest level, since the basic elements of the model, the informons, are generated using only causally local information. To do so does require abandoning the principle of continuity, at least at the lowest level, but continuity is regained as an emergent property via interpolation. Processes do act non-locally in the selection of informons to generate, but this does not involve information transfer. Non-locality in the form of outcome dependence in measurements is seen to be emergent, as is the probability structure of NRQM. For this reason the model is described as being quasi-local: local at the lowest level but non-local at emergent levels. The key point is that it is not locality which must be abandoned at the lowest level but rather the idea of independence of physical entities. The process approach is all about interactions among processes, and certain types of interactions, namely the interactive sums and products, involve correlations between the actions of the conjoined processes which destroys statistical independence. Reality can be local but the processes generating reality need not be independent. This is in accord with experience in psychology and biology.

The basic elements of the model are accorded definite properties, although not a complete set, due to the incompatibility of the sequential application of processes, akin to the non-commutativity of Hilbert space operators. Contextuality is a reflection of the fact that all measurement is a form of interaction between processes, and such interactions necessarily create conditions under which processes become active or inactive or combine or transform to form new processes. The process model thus asserts the existence of a definite reality whose elements possess definite properties, but these properties may be altered as a result of interaction with a measurement device. For this reason the model is described as quasi-non-contextual: primitive entities possess a definite, yet incomplete, set of properties which may only be observed through an act of measurement, which may alter them, and thus they are contextual at the emergent level. The observer, however, does not make the properties real. Process makes the properties real. The key insight is that processes interact and measurement is merely a specialized kind of process and so some measurements will alter the generating process of system while other measurements will not. This too is in accord with the view of psychology and biology, which accepts the presence of a definite external reality but that we cannot interact with that reality without sometimes altering it. 

The potential of interactive couplings of processes destroys the independence of processes and gives rise to an emergent non-Kolmogorov probability structure.  

Standard NRQM is seen, from the process perspective, to be an idealization. The finiteness of the process model renders it potentially testable, at least in principle. For finite values of $N,r$ there will be discrepancies between the Hilbert space interpretation generated via interpolation and the wave function obtained via Schr\"odinger theory. These arise partly due to the finite number of terms in the interpolation series and partly due to insufficient contributions to the strength causing it to fail to approximate the kernel. Another set of possible discrepancies arise when the Schr\"odinger wave function is not band limited relative to the choices of $t_{P},l_{P}$. If $t_{P},l_{P}$ are fixed and universal, say the Planck values or greater, then as the energy and momentum of a particle increases one might see discrepancies between the observed and theoretical wave function values. It might also be the case that as the energy increases it no longer becomes possible for a Schr\"odinger type wave function to be approximated and the processes generating the particle would need to decompose into lower energy and momenta. Observationally this would set an upper limit on the masses of particles. Breaking such limits would in turn place upper bounds on $t_{P}$ and $l_{P}$. I expect that other points of testability could be found.

The path integral strategy described above is but one of a large number of possible strategies capable of yielding NRQM in the asymptotic limit. It was presented here for its simplicity and clarity, not for its verisimilitude. A more accurate strategy requires the use of non-uniform interpolation techniques, generally involving functions other than sinc functions, but that requires mathematical techniques beyond the scope of the preset paper \cite{Zayed}. Sinc interpolation   works well for the case of scalar particles. It should extend fairly easily to that of vector particles although the interpolation theory becomes more complicated, especially when the components are strongly coupled. The current state of the art in interpolation theory does not extend very far into dealing with spinors but that may change in time. At present there is no a priori reason to believe that the general process approach will not apply to these more general particles, although the level of accuracy of the model may decline. This may stimulate interest in further expanding the range of interpolation theory.

Representing processes as combinatorial games opens up a rich mathematical field with lots of opportunity for research. The study of game strategies is particularly rich, and can be constrained by the requirement that they should result in weak epistemic equivalent models that satisfy at least the Schr\"odinger equation, if not perhaps many others. In fact the model shows that the Schr\"odinger equation may be viewed in emergent terms as well, being the asymptotic version of a local wave-like diffusion process.

The process model offers a new ontology for NRQM which is, to my mind at least, intuitive, comprehensible and involves fewer extreme assumptions. It offers an approach to NRQM which appears to be paradox free (or at lesat avoids currently known paradoxes) all the while permitting definite calculations to be carried out.

As an aside, the process approach has applications outside of quantum mechanics in the study of neural information flow and neurodynamics, functional differentiation and diversity, dynamic networks, gene regulation and expression, economics and so is worthy of study in its own right.

\begin{acknowledgments}
Thanks are due to Irina Trofimova and Robert Mann for many fruitful discussions.
\end{acknowledgments}

\end{document}